\documentclass[entropy,review,submit,pdftex,moreauthors]{Definitions/mdpi} 

\firstpage{1} 
\pubvolume{1}
\issuenum{1}
\articlenumber{0}
\pubyear{2024}
\copyrightyear{2024}
\datereceived{ } 
\dateaccepted{ } 
\datepublished{ } 
\hreflink{https://doi.org/} 
\pdfoutput=1 

\usepackage{physics}
\usepackage{subcaption}

\Title{Strongly Coupled $\mathcal{PT}$-Symmetric Models in Holography}

\TitleCitation{Strongly Coupled PT-Symmetric Models in Holography}

\Author{Daniel Areán $^{1,2}$*\orcidA{}, David Garcia-Fariña $^{1,2}$\orcidB{} and Karl Landsteiner $^{1}$\orcidC{}}

\AuthorNames{Daniel Areán, David Garcia-Fariña and Karl Landsteiner}

\AuthorCitation{Arean, D.; Garcia-Fariña, D.; Landsteiner, K.}

\address{%
${}^1$ \quad Instituto de F\'isica Te\'orica UAM/CSIC, c/Nicol\'as Cabrera 13-15, Universidad Aut\'onoma de Madrid, Cantoblanco, 28049 Madrid, Spain\\
${}^2$ \quad Departamento de F\'isica Te\'orica, Universidad Aut{\'o}noma de Madrid, Campus de Cantoblanco, 28049 Madrid, Spain}

\corres{Correspondence: daniel.arean@uam.es}

\abstract{Non-Hermitian quantum field theories are a promising tool to study open quantum systems.
These theories preserve unitarity if $\mathcal{PT}$-symmetry is respected, and in that case an equivalent Hermitian description exists via the so-called Dyson map.
Generically, $\mathcal{PT}$-symmetric non-Hermitian theories can also feature phases where $\mathcal{PT}$-symmetry is broken and unitarity is lost.
We review the construction of holographic duals to strongly coupled $\mathcal{PT}$-symmetric quantum field theories and the study of their phase diagram. 
We next focus on spacetime-dependent non-Hermitian couplings: non-Hermitian quenches and lattices. They violate the null energy condition in the gravity dual.
The lattices realize phases supporting an imaginary current that breaks $\mathcal{PT}$-symmetry spontaneously. Remarkably, these non-Hermitian lattices flow
to a $\mathcal{PT}$-symmetric fixed point in the IR.}


\begin{document}

\tableofcontents

\section{Introduction}

One of the fundamental axioms of Quantum Mechanics (QM) is that the Hamiltonian 
should be Hermitian. This assumption ensures unitary evolution or, equivalently, reality of the eigenvalues of the Hamiltonian. However, when taking unitarity as the true physical restriction, one concludes that it is possible to consistently formulate Quantum Mechanics with non-Hermitian Hamiltonians \cite{Bender:1998ke}. These non-Hermitian theories are instead required to have $\mathcal{PT}$ symmetry \cite{Bender:1998ke} or, more generally, any antilinear symmetry \cite{Mostafazadeh:2002id,Mannheim2018}. However, the existence of such symmetry is not a sufficient condition for unitary evolution as it can be spontaneously broken by the spectrum of the theory. These realizations gave rise to the field of $\mathcal{PT}$-symmetric Quantum Mechanics (see, for instance 
\cite{Bender_2005,bender:2019,Mostafazadeh:2020mdm,Ashida:2020dkc,Fring:2022tll} for an introduction).

Physically, a non-Hermitian theory is interpreted as having influx and outflux of matter. Thus, $\mathcal{PT}$-symmetric theories correspond to the family of non-Hermitian theories where influx and outflux are balanced, ensuring that matter content is conserved.
To illustrate these features, we consider the following $\mathcal{PT}$-symmetric two-level Hamiltonian \cite{Bender_2005}
\begin{equation}\label{eq:PTHamiltonian_2LevelSystem}
    H=\begin{pmatrix}-i \Gamma & g \\ g & i\Gamma\end{pmatrix}\,,
\end{equation}
where $\Gamma$ and $g$ are real. In this theory, $\mathcal{PT}$ is given by
\begin{equation}
    \mathcal{PT}=\begin{pmatrix}0  & 1 \\ 1 & 0\end{pmatrix}\mathcal{C}\,,
\end{equation}
with $\mathcal{C}$ the complex conjugation operator. Notably here, $\mathcal{P}$ and $\mathcal{T}$ have the usual interpretation of parity and time reversal.

The eigenvalues and eigenstates of the Hamiltonian $\eqref{eq:PTHamiltonian_2LevelSystem}$ are given by
\begin{align}
    &\psi_+=\frac{-i\Gamma+\sqrt{g^2-\Gamma^2}}{g} \begin{pmatrix}1 \\ 0 \end{pmatrix}+\begin{pmatrix}0 \\ 1 \end{pmatrix}\,,  \qquad H\psi_+=\sqrt{g^2-\Gamma^2}\,\psi_+\,,\nonumber\\
    &\psi_-= -\frac{i\Gamma+\sqrt{g^2-\Gamma^2}}{g} \begin{pmatrix}1 \\ 0 \end{pmatrix}+\begin{pmatrix}0 \\ 1 \end{pmatrix}\,, \qquad H\psi_-=-\sqrt{g^2-\Gamma^2}\,\psi_-\,.
\end{align}
When $g^2-\Gamma^2>0$, the eigenvalues are real and $\psi_+$ and $\psi_-$ are eigenstates of $\mathcal{PT}$
\begin{equation}
    \mathcal{PT}\psi_\pm = \frac{i\Gamma\pm\sqrt{g^2-\Gamma^2}}{g} \psi_\pm\nonumber\,.
\end{equation}
However, when $g^2-\Gamma^2<0$, the energies become complex and $\psi_+$ and $\psi_-$ are no longer eigenstates of $\mathcal{PT}$
\begin{equation}
    \mathcal{PT}\psi_\pm = \pm i\frac{\Gamma \mp \sqrt{\Gamma^2-g^2}}{g} \psi_\mp\nonumber\,.
\end{equation}
Hence the theory is only $\mathcal{PT}$-symmetric for $g^2-\Gamma^2>0$; which indeed matches the regime where the energies are real, showcasing the aforementioned relation between unitary evolution and $\mathcal{PT}$-symmetry. The value $g^2=\Gamma^2$, separating $\mathcal{PT}$-broken and $\mathcal{PT}$-unbroken regimes is labeled exceptional point.

The physical interpretation for this model is the following: $\Gamma$ encodes the transfer of matter between the exterior and the two levels and $g$ represents a hopping between them. Then, the $\mathcal{PT}$-symmetric regime corresponds to the hopping between levels being faster than the inflow (outflow) of matter from (to) the exterior. Hence the system is capable of finding equilibrium. Conversely, in the $\mathcal{PT}$-broken regime the inflow (outflow) is faster than the hopping and thus no stationary state is reached.

A particularly important feature of $\mathcal{PT}$-symmetric models is the existence of the Dyson map, a similarity transformation taking a non-Hermitian model with unbroken $\mathcal{PT}$-symmetry to a Hermitian theory. Such map allows $\mathcal{PT}$-symmetric models to be studied in terms of a more conventional Hermitian description. Nonetheless, in many cases, the mapping can be highly non-trivial and give rise to non-local interactions. This alone motivates the study of non-Hermitian theories for their own physical significance. Furthermore, once we remove the constraints of Hermiticity and instead construct $\mathcal{PT}$-symmetric Hamiltonians, it is interesting to study the vacua and the spontaneous breaking of $\mathcal{PT}$ as a function of the parameters of the model.  

Since the inception of $\mathcal{PT}$-symmetric quantum mechanics (QM), significant efforts have been made to extend the framework to quantum field theory (QFT) \cite{Bender_2004}. 
However, when studying $\mathcal{PT}$-symmetric QFTs, one faces a significant complication absent in QM: most QFTs cannot be solved non-perturbatively. For this reason, many of the studies 
of  $\mathcal{PT}$-symmetric QFTs have been restricted to the weak coupling limit. To name a few, 
in \cite{Mannheim:2018dur,Alexandre:2018xyy,Fring:2019xgw,Alexandre:2019jdb,Alexandre:2018uol,Fring:2019hue,Fring:2020bvr} the authors considered spontaneous symmetry breaking of global and gauge symmetries in $\mathcal{PT}$-symmetric QFTs, and 
in \cite{Chernodub:2021waz,Chernodub:2024lkr} the effects of non-Hermitian spacetime-dependent couplings were explored. 
However, the perturbative nature of these studies leads us to pose the following question: Does the phenomenology change significantly in the strong coupling regime? 
Gauge/gravity duality (also known as holographic duality or AdS/CFT correspondence) \cite{Maldacena:1997re, Aharony:1999ti,Ammon:2015,Hartnoll:2018xxg,zaanen2015holographic} offers a promising new avenue to find an answer.

In essence, gauge/gravity duality relates the large-$N$, strong coupling limit of certain (holographic) QFTs to the classical limit of a gravity theory in an asymptotically Anti-de Sitter (AdS) spacetime with an extra dimension, which encodes the renormalization group flow of the QFT. Thus, by solving problems of classical gravity, we can obtain non-perturbative results for a large family of models which can serve as candidates to probe the phenomenology of strongly coupled $\mathcal{PT}$-symmetric QFTs.
In this context, the standard approach, pioneered in \cite{Arean:2019pom}, is to consider a Hermitian holographic QFT and promote it to a $\mathcal{PT}$-symmetric one by introducing in the action a non-Hermitian coupling to some scalar operator $\mathcal{O}$ of the form
\begin{equation}
    \mathcal{S}_{\mathcal{O}}=\int d^dx\left(\bar{s}\mathcal{O}+s\bar{\mathcal{O}}\right)
\end{equation}
with $s$, $\bar{s}\in\mathbb{R}$ couplings (sources) satisfying $\bar{s}\neq s$. By tuning the values of $s$ and $\bar{s}$, one can probe different regimes of the theory, finding phases where $\mathcal{PT}$ is spontaneously broken and phases where it is not. Remarkably, this kind of construction is quite similar to the one employed in perturbative studies, where non-Hermiticities are also typically introduced through the couplings of the QFT.

The existing literature on $\mathcal{PT}$-symmetric holographic QFTs has been concerned with exploring and generalizing the model introduced in \cite{Arean:2019pom}. There the authors considered a holographic QFT in 3 spacetime dimensions and introduced the non-Hermiticity through the (constant) source of an operator $\mathcal{O}$ of conformal dimension $\Delta=2$ charged under a global $U(1)$ symmetry.
In \cite{Xian:2023zgu} the full phase diagram of this model and its transport properties were derived both at zero and non-zero chemical potential. The authors found phases where $\mathcal{PT}$ was spontaneously broken which were dual to a complex geometry in AdS and, remarkably, they concluded that the Ferrel-Glover-Tinkham (FGT) sum rule for the electric conductivity held for all phases (even those breaking $\mathcal{PT}$). Generalizations to spacetime-dependent sources were studied in \cite{Morales-Tejera:2022hyq} and \cite{Arean:2024lzz}. The former paper considered non-Hermitian quenches where the source was time-dependent and interpolated between different non-Hermitian theories. There the authors found that the null energy condition was violated. It was also emphasized that the Dyson map 
of the QFT \cite{Morales-Tejera:2022hyq} was dual to a gauge transformation in the gravity theory. On the other hand,
\cite{Arean:2024lzz} considered non-Hermitian lattices and junctions where the sources were dependent on one of the spatial coordinates. There it was observed that a non-Hermitian complex current appeared in the system concentrated around the regions where the derivative of the sources was greater. Interestingly, such current spontaneously breaks $\mathcal{PT}$ without the need for complex geometries. 
However, it was found that $\mathcal{PT}$-symmetry is recovered in the infrared (IR) of the theory. The inhomogeneous model~\cite{Arean:2024lzz} flows from a $\mathcal{PT}$-broken UV to a $\mathcal{PT}$-unbroken IR; similar to the results observed in \cite{Chernodub:2021waz} in the perturbative regime.

In this work we review recent progress in the formulation and study of $\mathcal{PT}$-symmetric holographic QFTs.
We focus mostly on the underlying ideas behind the construction
of these models and give an outline of 
the most interesting results. The paper is organized as follows. In section \ref{sect:GaugeGravity} we give a brief introduction to gauge/gravity duality. In section \ref{sect:BottomUpModel} we discuss the bottom-up construction of the holographic model introduced in \cite{Arean:2019pom}. There we follow closely the discussion presented in \cite{Xian:2023zgu} and \cite{Arean:2024lzz}. Following this, in section \ref{sect:ConstantCouplings} we present the full phase diagram and the transport properties of the model for constant coupling at zero and non-zero chemical potential. In section \ref{sect:spacetimesources} we discuss spacetime-dependent couplings, first in the context of non-Hermitian quenches and then in the context of non-Hermitian lattices. 
We briefly recap the main results in section \ref{sect:Conclusions} and  
conclude with section \ref{sect:FutureDirections} where we present a list of future directions and open questions.

\section{Gauge/gravity duality}\label{sect:GaugeGravity}
Here we provide a very brief heuristic introduction to gauge/gravity duality. We focus on introducing the key ingredients needed for the following sections. For an in-depth review of gauge/gravity duality and its applications we refer the reader to \cite{Ammon:2015,Hartnoll:2018xxg,zaanen2015holographic}.

Gauge/gravity duality establishes a relation between a theory of quantum gravity in a $(d+1)$-dimensional asymptotically AdS spacetime and a non-gravitational QFT in $d$ dimensions. Explicitly it posits that the generating functionals in both theories are equivalent
\begin{equation}
    \mathcal{Z}_{\text{QFT}}[J]=\mathcal{Z}_{\text{grav}}[\phi_0=J]
\end{equation}
where the sources $J$ in the QFT side are identified with the boundary conditions $\phi_0$ on the fields $\phi$ on the conformal boundary of AdS. Remarkably, gauge/gravity is a strong/weak duality. In the limit of strong coupling and large rank of the gauge group in the QFT side, the gravity dual is weakly coupled and the classical solution dominates 
\begin{equation}
    \mathcal{Z}_{\text{QFT}_d}[J]=\mathcal{Z}_{\text{grav}_{d+1}}[\phi_0=J]\approx e^{\mathcal{S}_{\text{grav}}^{\text{on-shell}}[\phi_0=J]}
\end{equation}
where $\mathcal{S}_{\text{grav}}^{\text{on-shell}}$ is the gravitational action evaluated on the classical solution. Hence, we reduce computations in a strongly-coupled QFT to solving classical equations of motion of gravity with adequate boundary conditions. In particular, correlation functions in the QFT can be computed by taking functional derivatives with respect to $\phi_0$
\begin{equation}\label{eq:n-point functions AdS/CFT}
    \expval{\mathcal{O}(x_1)...\mathcal{O}(x_n)}=\frac{\delta^n \mathcal{S}_{\text{grav}}^{\text{on-shell}}}{\delta\phi_0(x_1)...\,\delta\phi_0(x_n)}
\end{equation}
where $\mathcal{O}$ is the operator sourced by $J$, dual to the field $\phi$ in the gravity side. 

To make these ideas more transparent, it is convenient to consider AdS spacetime written in coordinates such that near the conformal boundary ($z=0$) the metric takes the form 
\begin{equation}
    ds^2 \approx \frac{l^2}{z^2}(-dt^2+d\Vec{x}^2+dz^2)
\end{equation}
where $l$ is the AdS scale which we set to 1 for convenience. 

The extra dimension $z$ is identified with the energy scale on the QFT. The UV corresponds to $z=0$ while the dynamics deep in the bulk ($z\gg0$) describe the IR. In fact, one thinks of the gravity theory as a geometrical realization of the renormalization group flow of the QFT. The flow from the UV towards the IR caused by the addition of a source $J$ to 
some relevant operator, is encoded in the flow from an AdS spacetime at $z\rightarrow0$ to a different geometry for $z\gg0$,
as a result of the boundary condition $\phi_0=J$. 

An important property of the duality is that 
the isometries of the 
spacetime are identified with symmetries of the QFT. 
In particular this implies that the UV QFT has to be invariant under the isometries of AdS, which are those of the conformal group in $d$ dimensions. Therefore the UV QFT is a conformal field theory (CFT).
Nonetheless, the isometry group of AdS typically is an isometry of only the $z\rightarrow0$ region; while the isometries deep in the bulk differ. This indicates that, generically, in gauge/gravity we study flows from a CFT in the UV to some IR QFT which need not be conformal. These flows are typically associated with the addition of some scale to the CFT, for instance by taking non-zero temperature or by sourcing some relevant operator.

The holographic dual to a CFT at finite temperature $T$ is given by an AdS black brane with temperature $T$.
That is, an asymptotically AdS spacetime with a planar horizon whose Hawking temperature is $T$.
On the other hand, sources have to be added trough boundary conditions of the fields on the AdS spacetime. To make this explicit, let us consider the asymptotic expansion of a generic bosonic field $\Phi$ near the conformal boundary $z=0$. The equations of motion are second order in $\partial_z$ and hence the solution is parameterized by two modes, a leading one $\phi_0$ and a subleading one $\phi_1$
\begin{equation}
    \Phi\approx \phi_0\, z^{\Delta_-}(1+...)+\phi_1\, z^{\Delta_+}(1+...)
\end{equation}
The leading mode $\phi_0$ is identified with the source $J$ of the dual operator $\mathcal{O}$ in the QFT. This dual operator has the same tensor structure as the field $\Phi$ and has conformal dimension $\Delta_+$. Using the relation \eqref{eq:n-point functions AdS/CFT} it can be seen that the vacuum expectation value (VEV) of $\mathcal{O}$ is proportional to the subleading mode $\phi_1$. The explicit values of $\Delta_\pm$ depend on the field $\Phi$; for instance, for a scalar field with mass $m$ living in $(d+1)$ dimensions we have $\Delta_\pm=d/2\pm\sqrt{d^2/4+m^2}$. We note that negative masses are allowed provided they are above the Breitenlohner-Freedman (BF) bound, which for the aforementioned scalar is given by $m^2>-d^2/4$.

Global symmetries and the corresponding conserved currents in the QFT are dual to gauge symmetries and the associated gauge fields in the gravity side, respectively. For instance, to consider
a $U(1)$ global symmetry we need to study the dynamics of a $U(1)$ gauge field $A_\mu$ in an asymptotically AdS spacetime. The gauge field will be dual to the $U(1)$ current operator $J_\mu$ and its equations of motion will yield the usual conservation law $\partial_\mu J^\mu=0$. The leading mode of $A_\mu$ is identified with an external gauge field which in the dual QFT would source $J_\mu$.

\section{Bottom-up holographic model}\label{sect:BottomUpModel}

A non-Hermitian holographic model was constructed in \cite{Arean:2019pom} and further developed in \cite{Morales-Tejera:2022hyq,Xian:2023zgu,Arean:2024lzz}. Let us now discuss how to construct it.
We follow a bottom-up approach, that is, we first consider the necessary ingredients in the QFT and then we construct the minimal gravity theory capable of reproducing those features.

From the point of view of the QFT, as stated in the introduction, we want to add the non-Hermicity through the source of an operator $\mathcal{O}$ of conformal dimension
$\Delta=2$ charged under a global $U(1)$ symmetry.
Hence the action we wish to study is given by
\begin{equation}\label{eq:actionNH_noCurrent}
    \mathcal{S}=\mathcal{S}_\text{CFT}+\int d^3x\left(\bar{s}\,\mathcal{O}+s\,\bar{\mathcal{O}}\right)
\end{equation}
where $\mathcal{S}_\text{CFT}$ is the Hermitian action, which we assume to be $\mathcal{PT}$-symmetric; and we have taken the QFT to live in 3 spacetime dimensions to match the setup of \cite{Arean:2019pom}.

Defining the action of $\mathcal{PT}$ on the coordinates as
\begin{equation}
    x=(t,x^1,x^2)\xrightarrow{\mathcal{PT}}\mathcal{PT}x=(-t,-x^1,x^2)\,,
\end{equation}
and taking $\mathcal{O}$ to be a scalar under $\mathcal{PT}$
\begin{equation}
    \mathcal{O}(x)\xrightarrow{\mathcal{PT}}\mathcal{O}(\mathcal{PT}x)\,,\qquad \bar{\mathcal{O}}(x)\xrightarrow{\mathcal{PT}}\bar{\mathcal{O}}(\mathcal{PT}x)\,,
\end{equation}
we thus conclude that the action \eqref{eq:actionNH_noCurrent} is invariant under $\mathcal{PT}$ as long as the sources $s$ and $\bar{s}$, which transform as functions,
\begin{equation}
    s(x)\xrightarrow{\mathcal{PT}}s^*(\mathcal{PT}x)\,,\qquad \bar{s}(x)\xrightarrow{\mathcal{PT}}\bar{s}^*(\mathcal{PT}x)\,,
\end{equation}
are real.

With this in mind, the minimal ingredients needed to reproduce such model in the gravity side are the following:
\begin{itemize}
    \item A $U(1)$ gauge symmetry dual to the global $U(1)$ symmetry.
    \item A complex scalar field $\phi$ of mass $m^2=-2$
    charged under the $U(1)$ gauge symmetry  dual to the operator $\mathcal{O}$. The leading order $\phi$ in the expansion around the conformal boundary of AdS is identified with the source $s$ (and correspondingly for $\bar{\phi}$).
\end{itemize}
Thus the minimal gravity action chosen in \cite{Arean:2019pom} is given by 
\begin{equation}\label{eq:Bulk_Grav_Action}
    \mathcal{S}=\int d^4y \sqrt{-g}\left(R-2\Lambda-\frac{1}{4}F_{MN}F^{MN}-\mathcal{D}_M\phi \mathcal{D}^M\bar{\phi}-m^2\phi\bar{\phi}-\frac{v}{2}\phi^2\bar{\phi}^2\right)
\end{equation}
where $A$ is the gauge field of the $U(1)$ symmetry, $F=dA$ is the field strength tensor, and $\Lambda=-3$ is the cosmological constant. The coupling $v=3$ is added so that at zero temperature the dual QFT has an RG flow interpolating between two conformal field theories \cite{Gubser:2008wz}. The action of the $U(1)$ transformation is 
\begin{equation}\label{eq:Bulk_GT}
    \phi\rightarrow e^{-iq\alpha}\phi\,,\qquad \bar{\phi}\rightarrow e^{iq\alpha}\bar{\phi}\,,\qquad A_M\rightarrow A_M-\partial_M\alpha\,,
\end{equation}
and the covariant derivatives are defined as 
\begin{equation}
    \mathcal{D}_M\phi=\partial_M\phi-iqA_M\phi\,,\qquad \mathcal{D}_M\bar{\phi}=\partial_M\bar{\phi}+iqA_M\bar{\phi}\,,
\end{equation}
where $q$ is the charge of $\mathcal{O}$, which we set to unity unless stated otherwise.

Remarkably, from the point of view of the gravity theory the non-Hermicity enters only through the boundary conditions on the conformal boundary ($z=0$) which explicitly read
\begin{align}\label{eq:UV_Boundary_Conditions}
    (z^2ds^2)|_{z\rightarrow 0}&=\left[- dt^2 +(dx^1)^2 + (dx^2)^2  + dz^2 \right]\,,\nonumber\\
    \partial_z\phi(z=0)=s\,,&\qquad \partial_z\bar{\phi}(z=0)=\bar{s}\,,\qquad A_\mu(z=0)=0\,.
\end{align}

It is worth noting that one can turn on sources for external gauge fields by modifying the boundary conditions of $A_\mu$. Concretely, if we take 
\begin{equation}
    A_\mu(z=0)=a_\mu
\end{equation}
this is dual to coupling the action \eqref{eq:actionNH_noCurrent} to an external gauge field $a_\mu$. In particular, the zero component of such gauge field $a_0$ can be interpreted as a chemical potential $\mu$. 

As reviewed in the introduction, $\mathcal{PT}$-symmetric models can be mapped to Hermitian theories through a similarity transformation known as Dyson map. Let us then discuss the implementation of the Dyson map in our holographic model as presented in \cite{Arean:2019pom} and further developed in \cite{Morales-Tejera:2022hyq,Arean:2024lzz}.
To do so, we focus first on the QFT side. We start from a Hermitian theory with $s=\bar{s}=M$ and consider the following similarity transformation (Dyson map)
\begin{equation}\label{eq:Similarity_operators}
    \mathcal{O}\rightarrow S\mathcal{O}\,,\qquad \bar{\mathcal{O}}\rightarrow S^{-1}\bar{\mathcal{O}}\,,
\end{equation}
where $S=e^{\beta(x)}$ is a complexified $U(1)$ transformation. At leading order, this transformation yields the following non-Hermitian action 
\begin{equation}\label{eq:Action_after_DysonMap}
    \mathcal{S}=\mathcal{S}_{\text{CFT}}+\int d^\text{3}x\,\,\left(MS\mathcal{O}+MS^{-1}\bar{\mathcal{O}}+\frac{i}{q}(S^{-1} \partial_\mu S)  J^\mu +O(S^{-1} \partial_\mu S)^2 \right)\,,
\end{equation}
where $J^\mu$ is the $U(1)$ current and $iq^{-1}S^{-1} \partial_\mu S$ behaves as an external gauge field. The term proportional to the current follows trivially from considering the variation of $\mathcal{S}_{\text{CFT}}$ under an infinitesimal $U(1)$ transformation
\begin{equation}
    \delta\mathcal{S}_{\text{CFT}}=\frac{i}{q}\int d^\text{3}x\,\,(S^{-1} \partial_\mu S)  J^\mu +O(S^{-1} \partial_\mu S)^2\,.
\end{equation}
Instead of considering the Dyson map \eqref{eq:Similarity_operators}, it is more convenient to reinterpret it by coupling the theory to an external gauge field $a_\mu$ and defining the Dyson map as the following external gauge transformation
\begin{equation}\label{eq:Dyson_Map}
   s\rightarrow sS^{-1}\,,\qquad \bar{s}\rightarrow \bar{s}S\,,\qquad a_\mu\rightarrow a_\mu+\frac{i}{q}S^{-1}\partial_\mu S\,,
\end{equation}
which also reproduces the action \eqref{eq:Action_after_DysonMap} if we start from a Hermitian theory with $s=\bar{s}=M$ and $a_\mu=0$.

From the point of view of the gravitational theory, the Dyson map \eqref{eq:Similarity_operators} is dual to a complexified gauge transformation \eqref{eq:Bulk_GT} with $S=e^{-i\alpha}$ and $\alpha=i\beta$. Remarkably, this changes the boundary conditions \eqref{eq:UV_Boundary_Conditions} to
\begin{align}\label{eq:UV_Boundary_Conditions_AfterDyson}
     (z^2ds^2)|_{z\rightarrow 0}&=\left[- dt^2 +(dx^1)^2 + (dx^2)^2  + dz^2 \right]\,,\nonumber\\
    \partial_z\phi(z=0)=S^{-1}s\,,&\qquad \partial_z\bar{\phi}(z=0)=S\bar{s}\,,\qquad A_\mu(z=0)=\frac{i}{q}S^{-1}\partial_\mu S\,.
\end{align}
where we have assumed that $\beta$ does not depend on $z$. Hence, in the gravity side we can also consider the alternative description of the Dyson map \eqref{eq:Dyson_Map} where we interpret it as an external gauge transformation acting on the sources
$s$, $\bar{s}$ and $a_\mu$.

Therefore, any theory
that under an external gauge transformation of the form \eqref{eq:Dyson_Map} can be mapped to a Hermitian one, admits a Hermitian description capable of reproducing the same phenomenological results.
We stress however that the Dyson map is not to be viewed as a gauge symmetry of the theory. Instead theories connected by Dyson maps describe different physical settings that share equivalent phenomenology \cite{Morales-Tejera:2022hyq}. Thus, following \cite{Arean:2019pom}, we choose to work with the sources $s$, $\bar{s}$ parametrized as
\begin{equation}\label{eq:SourcesParametrization}
    s=(1-\eta)M\,,\qquad \bar{s}=(1+\eta)M\,,
\end{equation}
where $\eta$ encodes the non-Hermiticity and $M$ is a dimensionful parameter.


\section{Phenomenology with constant sources}\label{sect:ConstantCouplings}

We begin now by reviewing some of the most relevant aspects of the model for constant sources. We focus on the phase diagram and the AC electric conductivity $\sigma(\omega)$ defined as
\begin{equation}
    \sigma(\omega)=\frac{\expval{J_1(\omega)}}{E_1(\omega)}\,,
\end{equation}
where $\expval{J_1}$ and $E_1$ are the $x^1$-components of the expectation value of the current and of the electric field, respectively. Note that we have used that the QFT is rotationally invariant in the $x^1-x^2$ plane to define a scalar conductivity as opposed to a tensorial one ($\sigma_{ij}=\delta_{ij}\sigma$). To compute this observable from the gravitational dual, one needs to consider linearized fluctuations of the gauge field $A_1$ (and of every other field that couples to it at the linear level).
This is equivalent to computing the conductivity in linear response theory in the QFT. For a background with rotational invariance in the $x^1-x^2$ plane, one needs to consider fluctuations of the metric $g_{\mu\nu}$ and the gauge field $A_\mu$ of the form
\begin{equation}
    \delta A_1=\alpha(z)e^{-i\omega t}\,,\qquad\delta g_{01}=h_{01}(z)e^{-i\omega t}
\end{equation}
where $z$ is the holographic coordinate. Then, the conductivity reads
\begin{equation}
    \sigma(\omega)=\frac{\partial_z\alpha(z=0)}{i\omega \alpha(z=0)}\,.
\end{equation}
Note that this is a direct application of the holographic dictionary as $J_1$ is the operator dual to $\delta A_1$ and $E_1$ is minus  the time derivative of the corresponding source. Hence $\expval{J_1(\omega)}=\partial_z\alpha(z=0)$ and $E_1(\omega) =i\omega \alpha(z=0)$.

For the metric in this section it is convenient to take the following ansatz in Poincaré coordinates
\begin{equation}
    ds^2=\frac{1}{z^2}\left(-u e^{-\chi}dt^2+(dx^1)^2+(dx^2)^2+\frac{dz^2}{u}\right)\,,
\end{equation}
where $u$ and $\chi$ are functions of the holographic coordinate $z$. 

We divide this section in two subsections, discussing solutions with zero and non-zero chemical potential; respectively.

\subsection{Phase diagram and conductivity at $\mu=0$}
\begin{figure}
    \centering
    \includegraphics[width=0.75\linewidth]{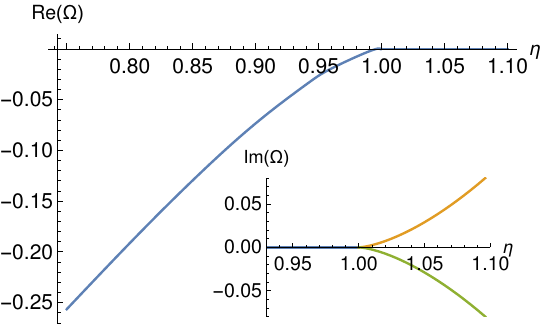}
    \caption{Free energy $\Omega$ for the model with constant sources at $T=0$ and $\mu=0$. 
    Note how for $|\eta|>1$ the free energy becomes complex thus signaling the spontaneous breaking of $\mathcal{PT}$. This figure has been adapted from \cite{Arean:2019pom}.}
    \label{fig:FreeEnergy_Arean}
\end{figure}

The phase diagram at zero chemical potential and zero temperature was first obtained in \cite{Arean:2019pom}. There the authors showed that the system presents the following phases depending on the value of $|\eta|$ (see figure \ref{fig:FreeEnergy_Arean}) 
\begin{description}
    \item[$\boldsymbol{|\eta|<1}$:] The QFT is in the $\mathcal{PT}$-symmetric phase and the free energy is real. In the gravitational dual the metric is real and the null energy condition (NEC) is satisfied. 
    \item[$\boldsymbol{|\eta|>1}$:] The QFT is in the $\mathcal{PT}$-broken phase. There are two branches of solutions whose free energies are complex conjugate to each other. In the gravitational dual the metric is complex and the NEC is ill-defined.
    \item[$\boldsymbol{|\eta|=1}$:] The QFT is at the exceptional point between the $\mathcal{PT}$-symmetric and the $\mathcal{PT}$-broken phases. In the gravitational dual the metric and the NEC is satisfied. Remarkably in this case the scalar field decouples from the metric and does not backreact.
\end{description}

\begin{figure}
\captionsetup[subfigure]{justification=centering}
    \begin{subfigure}{0.405\linewidth}
        \includegraphics[width=\linewidth]{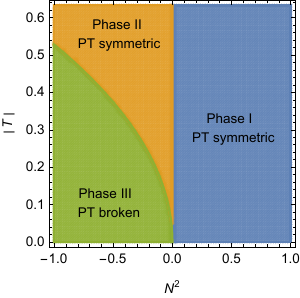}
        \caption{$\mu=0$}
        \label{fig:PhaseDiagram_mu0_Meyer}
    \end{subfigure}
    \begin{subfigure}{0.495\linewidth}
        \includegraphics[width=\linewidth]{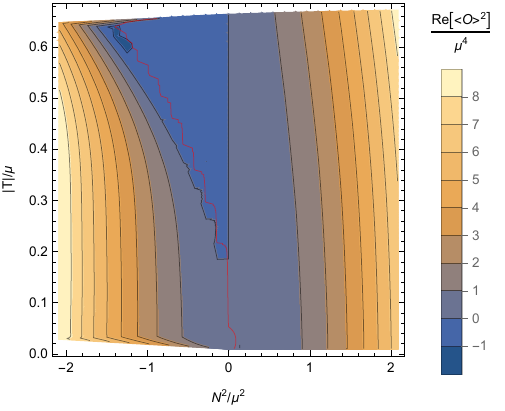}
        \caption{$\mu\neq0$}
        \label{fig:PhaseDiagram_muneq0_Meyer}
    \end{subfigure}
    \caption{Phase diagram for the model with constant sources at zero chemical potential (a) and non-zero chemical potential (b). Here $N^2=(1-\eta^2)M^2$. In (b) the color map represents the quantity $\Re[\expval{\mathcal{O}}^2]/\mu^4$. In figure (b) Phase I is found in the region with $N^2/\mu^2>0$  
    and the red curve denotes the transition between phase II (right) and phase III (left). The superconducting phase appears for $|T/\mu|<0.02$ and thus is not plotted. Both figures have been taken from \cite{Xian:2023zgu}.}
    \label{fig:PhaseDiagram_Meyer}
\end{figure}

The phase diagram at non-zero temperature was constructed in \cite{Xian:2023zgu} where the following phase structure was found (see figure \ref{fig:PhaseDiagram_mu0_Meyer})
\begin{description}
   \item[Phase I:] In the region with $|\eta|<1$, the QFT is in a $\mathcal{PT}$-symmetric phase and the free energy is real. In the gravitational dual the metric is real and the null energy condition (NEC) is satisfied. The zero-temperature limit of this phase corresponds to the phase with $|\eta|<1$ above.
   \item[Phase II:]  In the region with $1<|\eta|<\sqrt{1+\lambda_c (T/M)^2}$ with $\lambda_c\approx 3.6$,
   the QFT is in a $\mathcal{PT}$-symmetric phase and the free energy is real. In the gravitational dual the metric is real but the NEC is violated. This phase is linearly unstable. The zero-temperature limit of this phase corresponds to the exceptional point $|\eta|=1$ above.
   \item[Phase III:] In the region with $|\eta|>\sqrt{1+\lambda_c (T/M)^2}$,
   the QFT is in a $\mathcal{PT}$-broken phase. There are two branches of solutions whose free energies are complex conjugate to each other. In the gravitational dual the metric is complex, NEC is ill-defined, and the temperature is complex. This phase is linearly unstable. The zero-temperature limit of this phase corresponds to the phase with $|\eta|>1$ above. 
\end{description}

With regards to the conductivity, the authors of \cite{Xian:2023zgu} found that all phases satisfy the Ferrel-Glover-Tinkham (FGT) sum rule 
\begin{equation}
    \int_{-\infty}^\infty d\omega (Re[\sigma(\omega)]-1)=0\,.
\end{equation}
This sum rule is derived from assumptions of causality, unitarity and charge conservation; hence it is quite remarkable that it still holds in phase III where unitarity is violated due to $\mathcal{PT}$ being spontaneously broken.

\begin{figure}
\captionsetup[subfigure]{justification=centering}
    \centering
    \begin{subfigure}{0.32\linewidth}
        \includegraphics[width=\linewidth]{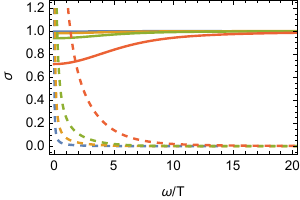}
        \caption{Phase I.}
        \label{fig:CondR1_Meyer}
    \end{subfigure}
    \begin{subfigure}{0.32\linewidth}
        \includegraphics[width=\linewidth]{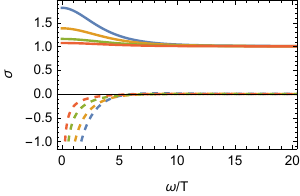}
        \caption{Phase II.}
        \label{fig:CondR2_Meyer}
    \end{subfigure}
    \begin{subfigure}{0.32\linewidth}
        \includegraphics[width=\linewidth]{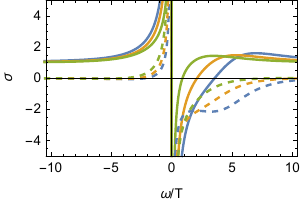}
        \caption{Phase III.}
        \label{fig:CondR3_Meyer}
    \end{subfigure}
    \caption{Conductivity $\sigma$ in the three different phases of the model with constant sources for $T\neq0$ and $\mu=0$. The real (imaginary) part is denoted by solid (dashed) lines. The colors \{blue, yellow, green, red\} correspond to solutions with $\expval{\mathcal{O}}/(NT)=\{-1.6,-1.7,-2.0,-3.5\}$ in (a) and to solutions with $\expval{\mathcal{O}}/(NT)=\{1.3,0,-0.9,-1.3\}$ in (b). In (c) The colors \{blue, yellow, green\} correspond to solutions with $N^2/|T|^2=\{-17.0,-8.8,-4.6\}$. Here $N^2=(1-\eta^2)M^2$. All figures have been taken from \cite{Xian:2023zgu}.}
    \label{fig:Cond_Meyer}
\end{figure}

The profile of the conductivity is shown in figure \ref{fig:Cond_Meyer} for the three different phases. As the QFT is conformal in the UV, the large frequency limit of the conductivity recovers the conformal result $\sigma=1$. 
At zero frequency the real part  of the conductivity displays a delta function typical of translational invariant systems. As the Krammers-Kronig relation is obeyed, the imaginary part features a $1/\omega$ pole and indeed the low frequency behavior of the conductivity is given by
\begin{equation}
    \sigma(\omega)=\rho\left(\pi\delta(\omega)+\frac{i}{\omega}\right)+...\,,
\end{equation}
where the ellipses
denote terms regular in $\omega$, and $\rho$ is the total charge density (e.g. for a superfluid $\rho$ would be the sum of the superfluid charge density and the normal charge density). In phase I, the charge density satisfies $\rho>0$ and the conductivity obeys $\sigma(\omega)^*=\sigma(-\omega)$ as one would expect in a Hermitian theory. On the other hand, in phase II the conductivity still fulfills $\sigma(\omega)^*=\sigma(-\omega)$ but the charge density is now negative $\rho<0$. Lastly, in phase III the charge density becomes complex $\rho\in\mathbb{C}$, and as $\mathcal{PT}$ is spontaneously broken $\sigma(\omega)^*\neq\sigma(-\omega)$.

\subsection{Phase diagram and conductivity at $\mu\neq0$}
Solutions with $\mu\neq0$ were studied in \cite{Xian:2023zgu}. In the presence of a chemical potential the model \eqref{eq:Bulk_Grav_Action} has a superconducting instability where the scalar field $\phi$ acquires a non-trivial profile even in the absence of sources~\cite{Gubser:2008px,Hartnoll:2008vx}. In the dual QFT, this corresponds to a phase transition from a normal fluid to a superfluid associated to the spontaneous symmetry breaking of the global $U(1)$ symmetry. For $|\eta|=1$ the phase transition is second order while for $|\eta|\neq1$ it is a crossover.\footnote{Properly speaking, the $U(1)$ symmetry is not spontaneously broken as it already was explicitly broken by the sources $s$ and $\bar{s}$. This in turn implies that the superconductor phase transition is generically expected to be a crossover as it is indeed observed.}

The phase diagram
can be seen in figure \ref{fig:PhaseDiagram_muneq0_Meyer}. Note that the superconducting phase only appears below $|T/\mu|\approx0.02$ and is not plotted in the figure. For temperatures above $|T/\mu|\approx0.02$ the phase structure matches the one found at zero chemical potential.

\begin{figure}
    \centering
    \includegraphics[width=0.75\linewidth]{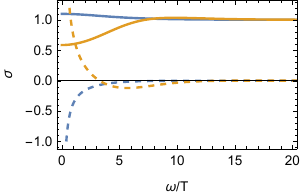}
    \caption{Conductivity $\sigma$ in phase II for the model with constant sources at $T\neq0$ and $\mu\neq0$. The real (imaginary) part is denoted by solid (dashed) lines. The blue (orange) line denotes a solution with $N^2/\mu^2=-6.2$ and $T/\mu$=2.4 ($N^2/\mu^2=-0.063$ and $T/\mu$=0.22). The change of sign in $\Im\sigma(\omega\rightarrow0)$ indicates a change in the sign of the charge density. Note how, at sufficiently small $|N^2/\mu^2|$, the charge density becomes positive as the normal component dominates over the negative contribution. Here $N^2=(1-\eta^2)M^2$. This figure has been taken from \cite{Xian:2023zgu}.}
    \label{fig:Cond_Charge_Meyer}
\end{figure}

With regards to the conductivity, once again the FGT sum rule holds for all phases. Moreover, the properties of $\sigma(\omega)$ are the same as those discussed in the previous section with only one minor difference in phase II. Now as the fluid has non-zero chemical potential the charge density gets a positive contribution from the normal charge density which can win against the negative contribution found in phase II
at zero chemical potential. This behaviour, illustrated in figure \ref{fig:Cond_Charge_Meyer},
 is only observed for sufficiently small values of the ratio $|(1-\eta^2)M^2/\mu^2|$.
 
\section{Phenomenology with spacetime-dependent sources}
\label{sect:spacetimesources}

In this section we will analyze holographic theories with spacetime-dependent non-Hermitian sources.  We first consider quenches of those couplings and next study lattices where the non-Hermitian sources are periodic along a spatial direction.

The study of quantum field theories with spacetime-dependent sources is 
compelling because it reveals insights into how quantum fields respond to external conditions that vary across space and time, expanding our understanding of fundamental physics. Studying non-Hermitian $\mathcal{PT}$-symmetric quantum field theories with spacetime-dependent sources is intriguing as these theories challenge and expand the fundamental principles of quantum mechanics and field theory.

We are interested in non-Hermitian quenches as they can be used to describe the rapid transition from a closed (Hermitian) system to an open (non-Hermitian) one.
Non-Hermitian lattices instead describe physical setups where an influx/outflux of matter 
is distributed periodically along the system.
They can also be useful to model a junction between a non-Hermitian and a Hermitian system.

\subsection{Non-Hermitian quenches}
\label{ssect:nhquenches}
Quantum quench refers to a sudden change in the Hamiltonian of a quantum system at or around some specific time. More generally we refer to a quench as a time dependence of the Hamiltonian localized in time around say $t=0$ and of a duration $\tau$. Thus for $|t|\gg\tau$ one is essentially dealing with a time independent Hamiltonian. It is of highest interest to see how the initial ground state evolves in time and eventually settles down to a new ground state. In quantum field theory such a problem is extremely difficult to address, especially  so in the strong coupling regime. It is rather remarkable that in gauge/gravity duality the problem is much easier to deal with. Such holographic quantum quenches have lead to important insights into questions of thermalization in strongly coupled systems (see e.g. \cite{Chesler:2013lia} for a review).

In the case of quantum mechanical systems with time-dependent pseudo-Hermitian Hamiltonians two very valuable
reviews are \cite{Mostafazadeh:2020mdm,Fring:2022tll}.
In gauge/gravity duality a completely analogue time dependence can be introduced in the sources 
at the boundary
of the asymptotically AdS space and this allows to study non-Hermitian quantum quenches of strongly coupled holographic quantum field theories.
In view of this it is rather exciting to apply the method of holographic quantum quenches to the non-Hermitian model introduced in the previous section. 
This program has been carried out in \cite{Morales-Tejera:2022hyq}.
We will now briefly review some of the most important results of this study.  
The holographic model~\eqref{eq:Bulk_Grav_Action} was solved in presence of time-dependent non-Hermitian sources. 
That is, we take a source configuration as in~\eqref{eq:SourcesParametrization} where now the non-Hermitian deformation is a time-dependent function $\eta=\eta(t)$.

In order to study the time-dependent problem it is helpful to use the following ansatz for the metric in terms of ingoing Eddington-Finkelstein coordinates
\begin{align}
    ds^2 = -f dv^2 + \frac{2}{z} e^{g} dz dv + h(dx^2+dy^2)\,,
\end{align}
where $f$, $g$, and $h$ are functions of the coordinates $(v,z)$.
We require the asymptotic boundary conditions
\begin{align}
    \lim_{z\rightarrow 0}(z^2f)=1~~~,~~~\lim_{z\rightarrow 0}(g)=0~~~,~~~\lim_{z\rightarrow 0}h=1\,,
\end{align}
that fix the asymptotic AdS boundary at $z=0$. Naturally we switch on the scalar fields $\phi$ and $\bar\phi$ as functions of $(v,z)$ as well, and, additionally, one has to allow for a nonzero time component of the gauge field $A_v(v,z)$. 
Moreover, since we are interested in finite temperature solutions, we require our geometries to feature a non-degenerate planar apparent horizon where $f=0$. Its coordinate location can be set to $z=1$. 

The equations of motion resulting from~\eqref{eq:Bulk_Grav_Action} are now partial differential equations (PDEs). They were solved numerically in~\cite{Morales-Tejera:2022hyq} where the time evolution of the following operators was monitored:
the scalar operators $\mathcal{O}$ and $\mathcal{\bar O}$
dual to the scalar fields $\phi$ and $\bar\phi$; the energy density $T_{vv}$ and pressure $T_{xx}=T_{yy}$ dual to the metric components $g_{vv}$ and $g_{xx}$, $g_{yy}$; and the charge density $J^v$ dual to the gauge field $A_v$. It is important to note that we do not source the gauge field explicitly and due to the non-Hermitian nature of the boundary conditions the gauge field turns out to be purely imaginary. Full details on the expression of these operators in terms of the AdS fields and on the numerical procedures can be found in \cite{Morales-Tejera:2022hyq}. 

\begin{figure}[h!]
    \centering
    \subfloat{\includegraphics[scale=0.32]{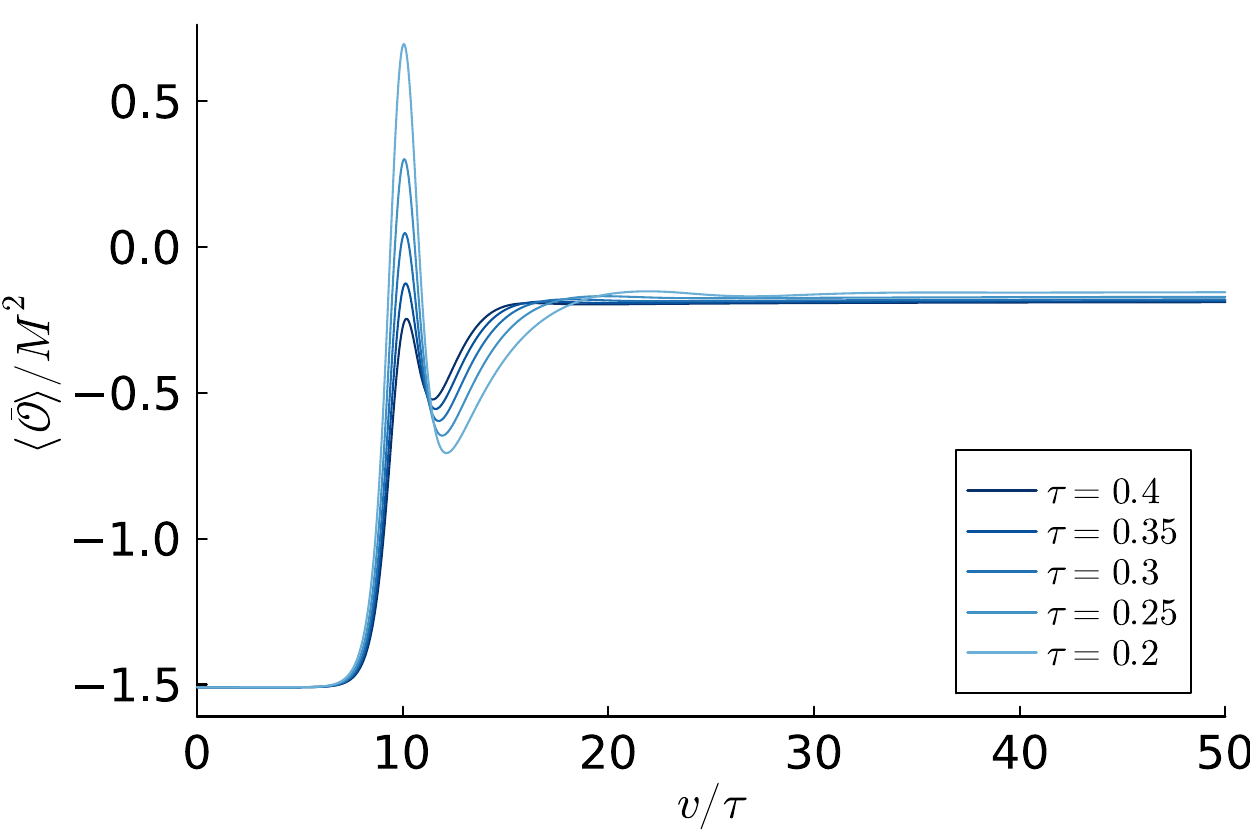}}
    \subfloat{\includegraphics[scale=0.32]{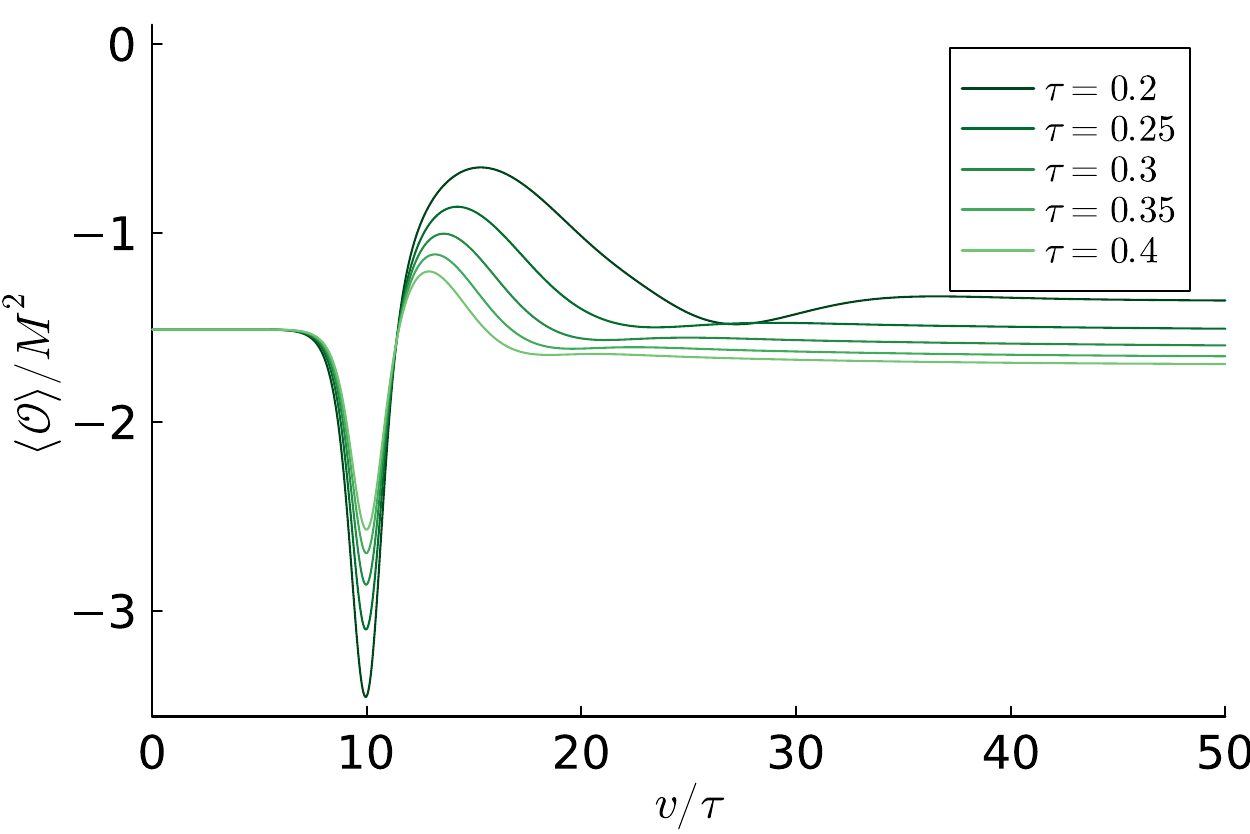}}
     \qquad
    \subfloat{\includegraphics[scale=0.32]{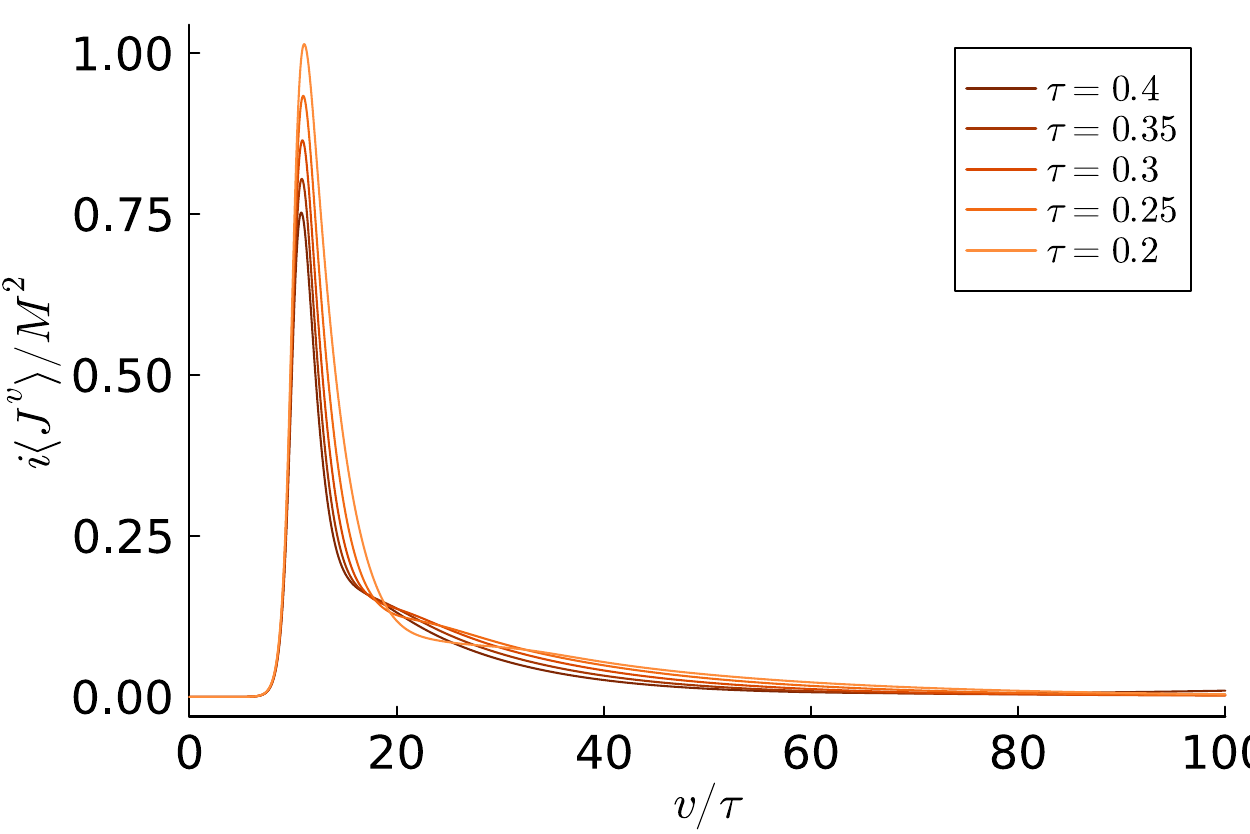}}
    \subfloat{\includegraphics[scale=0.32]{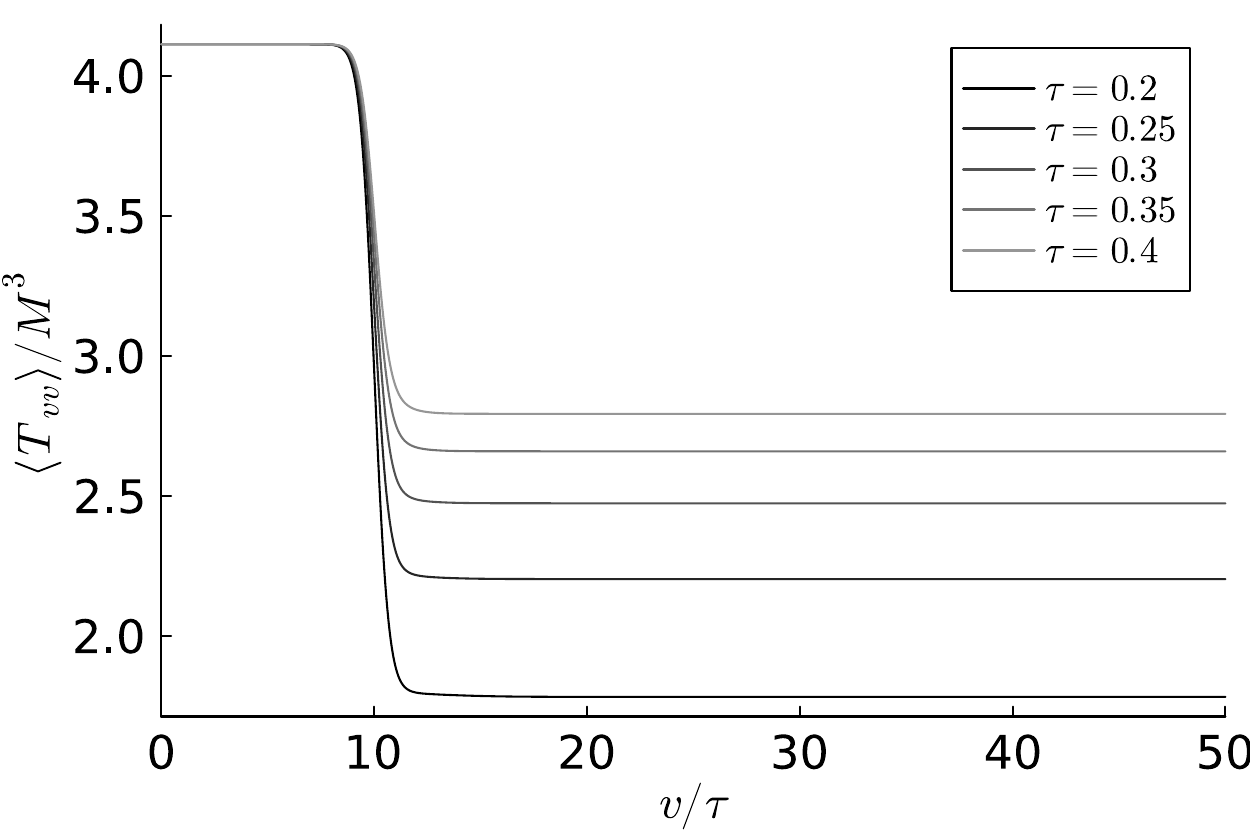}}
    \caption{Expectation values of scalar operators, (imaginary) charge density and energy density for a quench with profile \eqref{eq:quench} for several values of $\tau$ which interpolates between the Hermitian point $\eta_i=0$ and a final value $\eta_f=0.8\,$. (Adapted from \cite{Morales-Tejera:2022hyq})
    }
    \label{fig:quench1}
\end{figure}

The non-Hermitian (holographic) quantum quench is implemented by the following choice of profile for $\eta$
\begin{align}\label{eq:quench}
    \eta(v) = \eta_i+\frac{\eta_f-\eta_i}{2}\left[1 + \tanh\left( \frac{v-v_m}{\tau}\right)\right]\,.
\end{align}
The initial value $\eta_i$ is chosen to be the Hermitian theory $\eta_i=0$ whereas the endpoint is taken to be $\eta_f=0.8$. We remind the reader that the exceptional point at the phase transition towards the $\mathcal{PT}$-broken phase is $\eta=1$. The results of such quantum quenches with $v_m=10\tau$ for different values of $\tau$ are shown in figure \ref{fig:quench1}.

\begin{figure}
    \centering
    \subfloat{\includegraphics[scale=0.32]{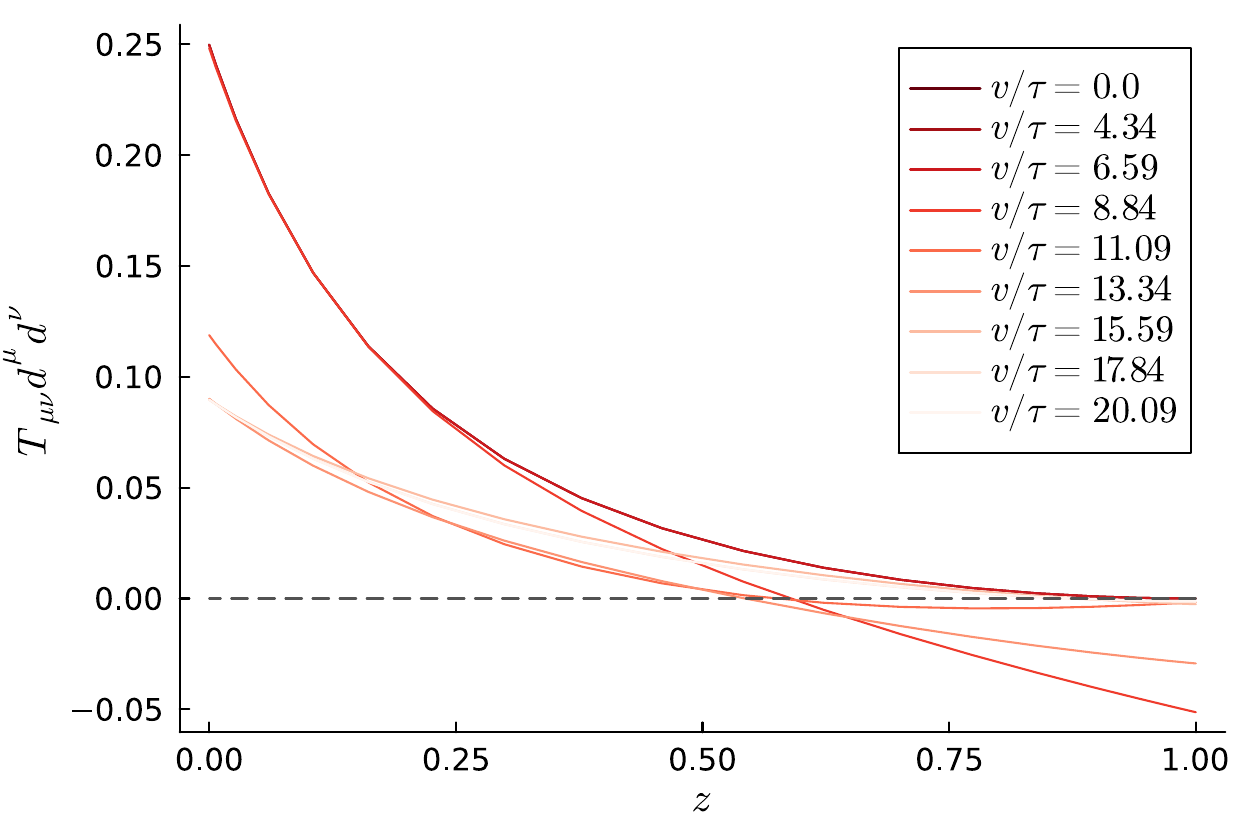}}
    \subfloat{\includegraphics[scale=0.32]{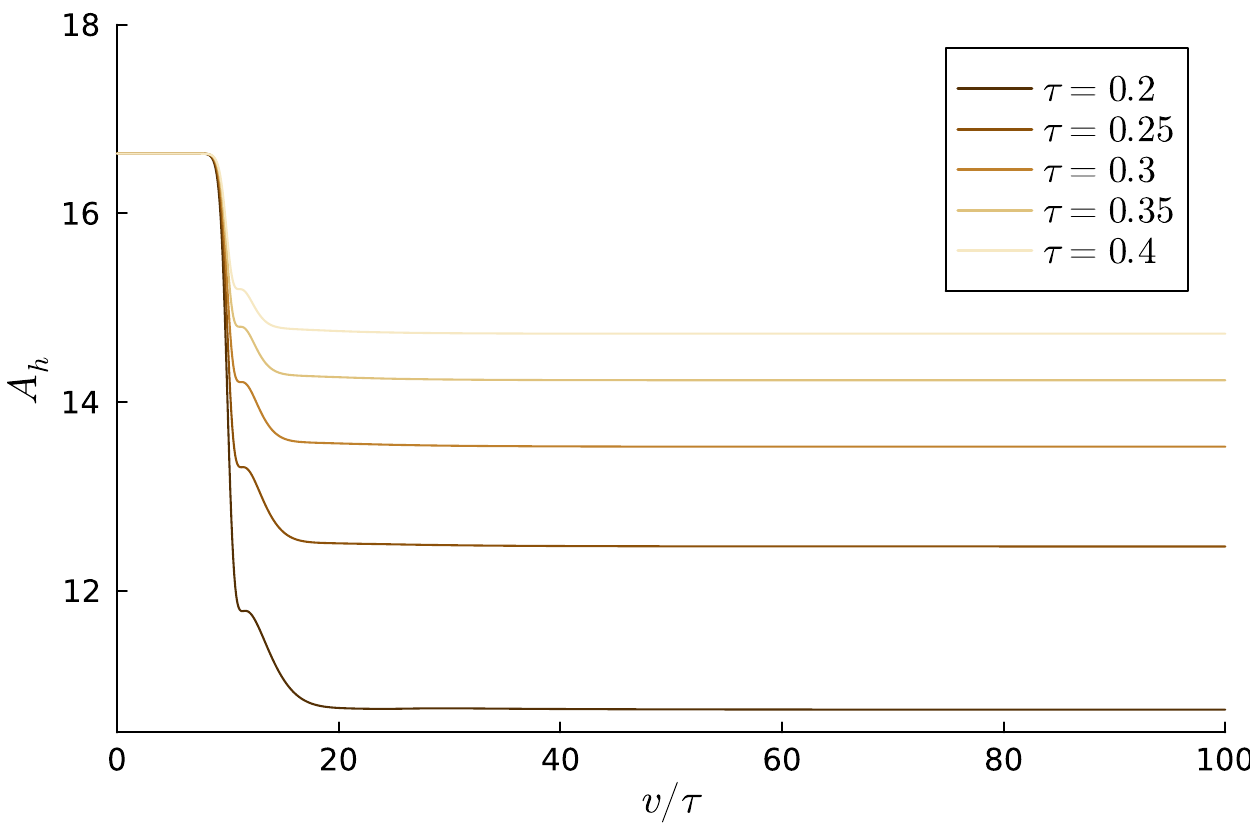}}
    \caption{Left: The Null Energy Condition $T_{\mu\nu}d^{\mu}d^{\nu}\geq 0$ for the particular simulation of figure \ref{fig:quench1} with $\tau=0.4\,$ at different stages $v/\tau$ of the evolution. It is clearly violated, especially at the horizon $z = 1\,$. 
    Right: We display the size of the apparent horizon for the simulations of figure \ref{fig:quench1}. All of them show a shrinking horizon, which is tightly related to violating the NEC. }
    \label{fig:violNEC}
\end{figure}

The most interesting feature is that the energy of the dual field theory decreases. Since the start and end points are black branes this means that the temperature of the black brane is actually lowered by the non-Hermitian quench. In fact the apparent horizon shrinks during the time evolution. This shrinking of the horizon is not in contradiction to the second law of black hole thermodynamics, which states that the area of the horizon has to grow in any given process provided the null energy condition (NEC) is satisfied. Indeed, during the non-Hermitian quench the NEC is violated. It suffices to take the null vector tangent to infalling null geodesics $d = \partial_v-\frac 1 2 z^2f e^{-g}\partial_z$. Then $T_{\mu\nu}d^{\mu}d^{\nu}$, where $T_{\mu\nu}$ is the bulk energy momentum tensor, should be positive if the NEC was to hold. The violation of the NEC can be clearly seen in figure \ref{fig:violNEC}

We note that \cite{Morales-Tejera:2022hyq} also investigated quenches that end in the exceptional point. There it was found that the endpoint $\eta_f=1$
settled down to new equilibrium configurations with non-vanishing charge density. 
Until now we have considered the scenario where $\eta$ corresponds to quenches in which a non-Hermitian operator is time-dependent. Instead one can study configurations
where the coupling of a Hermitian operator is quenched about a
$\mathcal{PT}$-symmetric point. 
Interestingly in this case the NEC is still violated in the bulk of the space time but not at the Horizon.
Consequently for such quenches the black brane horizon grows!

Finally, we could instead consider a non-Hermitian time-dependent configuration of sources equivalent via the Dyson map~\eqref{eq:Dyson_Map} to a Hermitian quench. It is clear from the form of ~\eqref{eq:Dyson_Map} that one needs to switch on a purely imaginary source for the gauge field $A_v$ to have such non-Hermitian configuration. Indeed in \cite{Morales-Tejera:2022hyq} it was checked that the time evolution of the holographic model with asymptotic boundary conditions
\begin{equation}
\label{eq:nHquenchsources}
\partial_z\phi(z=0,v) = (1-\eta)M\,,\qquad
\partial_z\bar \phi(z=0,v) = (1+\eta)M\,,\qquad
a_v(z=0,v) = \frac{i}{q} \frac{\partial_v\eta}{1-\eta^2}\,,
\end{equation}
results in the same dynamics as that obtained from imposing
\begin{equation}
\label{eq:dysonquenchsources}
\partial_z\phi(z=0,v) = \sqrt{1-\eta^2}M\,,\qquad
\partial_z\bar \phi(z=0,v) = \sqrt{1-\eta^2}M\,,\qquad
a_v(z=0,v) = 0\,,
\end{equation}
which is the source configuration resulting from applying the map~\eqref{eq:Dyson_Map} to
\eqref{eq:nHquenchsources} with $S^{-1}=\sqrt{(1+\eta)/(1-\eta)}$.

\subsection{Non-Hermitian lattices}
In this subsection we review the results of~\cite{Arean:2024lzz} and consider setups where the non-Hermitian deformation of the holographic theory is inhomogeneous, namely we make $\eta$ a function of a spatial coordinate. In particular, we take the sources
$s$ and $\bar{s}$ to be functions of the coordinate $x^1$.
In \cite{Arean:2024lzz} this allowed the authors to
build a \textit{non-Hermitian lattice} and a \textit{non-Hermitian junction.}
The former is a lattice where in each site there is an inhomogeneous (space-dependent) inflow/outflow of matter; and the latter is
a Hermitian/non-Hermitian/Hermitian junction which could describe a lattice system where in one site there is an impurity causing the inflow/outflow. For the sake of conciseness, here we focus only on the lattice case. In \cite{Arean:2024lzz}, this setup was studied in more detail due to its greater numerical simplicity. Similar conclusions are expected to hold for the junction.

The lattice is characterized by the following inhomogeneous non-Hermitian parameter 
\begin{equation}\label{eq:eta_Lattice}
    \eta(x^1)=a\cos\left(\frac{2\pi}{L} x^1\right)\,,
\end{equation}
and a vanishing external gauge field ($a_\mu=0$).\footnote{In \cite{Arean:2024lzz} the authors also considered non-vanishing $a_1=iq^{-1}\partial_1\eta/(1-\eta^2)$. This allowed them to explicitly check that, as pointed out in \cite{Morales-Tejera:2022hyq}, the non-Hermitian model is a Dyson map of the Hermitian theory.}
Due to the conformal symmetry of the UV the non-Hermitian lattices are thus characterized by 3 parameters, the temperature $T/M$, the cell length $LM$ and the amplitude $a$. 

The dual geometry is going to be inhomogeneous along the $x^1$ direction and can be solved by means of the following metric ansatz:
\begin{align}\label{eq:Ansatz_Poincare}
    ds^2=\frac{1}{z^2}&\left[- \left(1-z^3\right)h_1 dt^2 + h_3 \left(dx^1 + h_5\, dz\right)^2 +h_4 (dx^2)^2  + \frac{h_2}{1-z^3}  dz^2 \right]\,,
\end{align}
where $h_1$, ..., $h_5$ are functions of $(z,x)$. Note that this ansatz has residual gauge symmetries that have not been removed. Thus, the PDEs resulting from the equations of motion are solved numerically using the DeTurck trick \cite{Arean:2024lzz}.

\begin{figure}
\captionsetup[subfigure]{justification=centering}
\centering
\begin{subfigure}{\textwidth}
    \includegraphics[width=\textwidth]{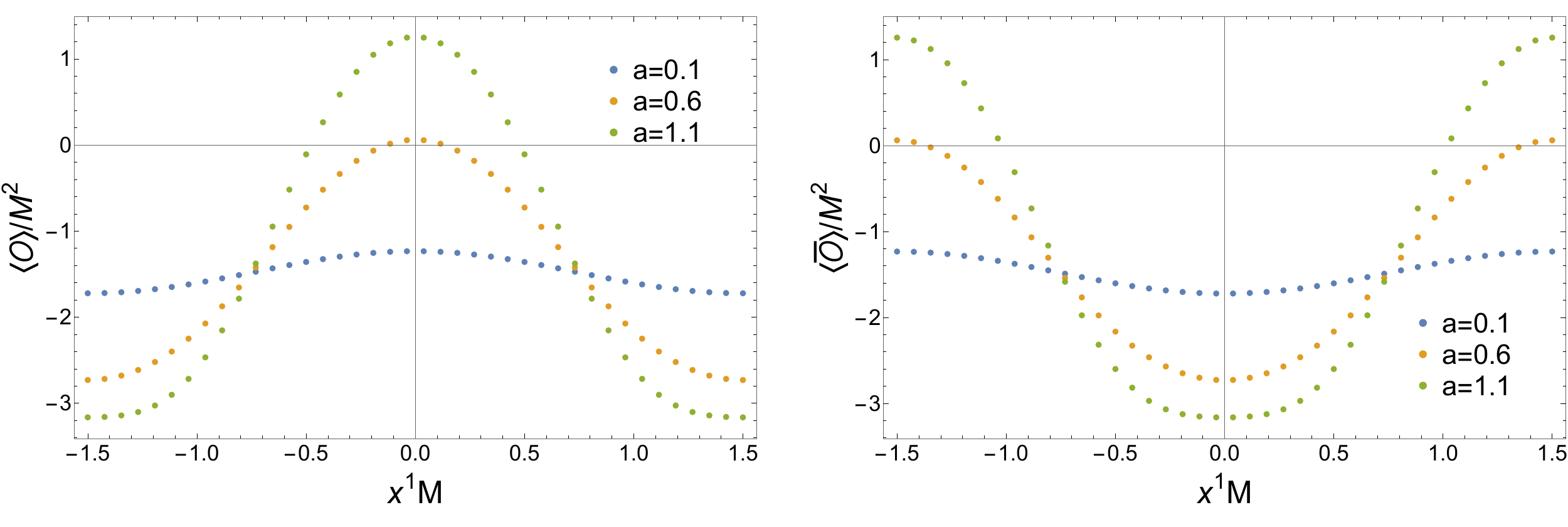}
    \caption{$T/M=0.5$}
    \label{fig:Os_PTbroken_T0d5}
\end{subfigure}
\hfill
\hfill
\begin{subfigure}{\textwidth}
    \includegraphics[width=\textwidth]{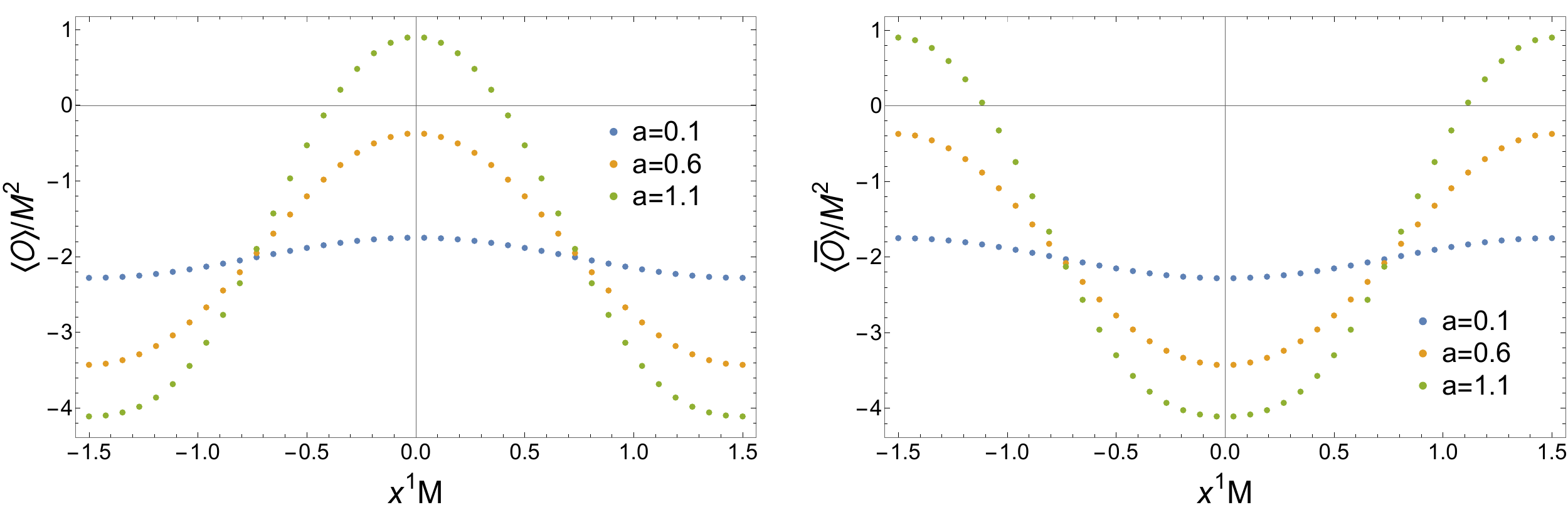}
    \caption{$T/M=1$}
    \label{fig:Os_PTbroken_T1d0}
\end{subfigure}        
\caption{Expectation values of the operators $\expval{{\mathcal{O}}}$ and $\expval{\bar{\mathcal{O}}}$ in a non-Hermitian lattice with $LM=3$. Figure taken from \cite{Arean:2024lzz}.}
\label{fig:Os_PTbroken}
\end{figure}

\begin{figure}
\captionsetup[subfigure]{justification=centering}
\centering
\begin{subfigure}{.49\textwidth}
    \includegraphics[width=\textwidth]{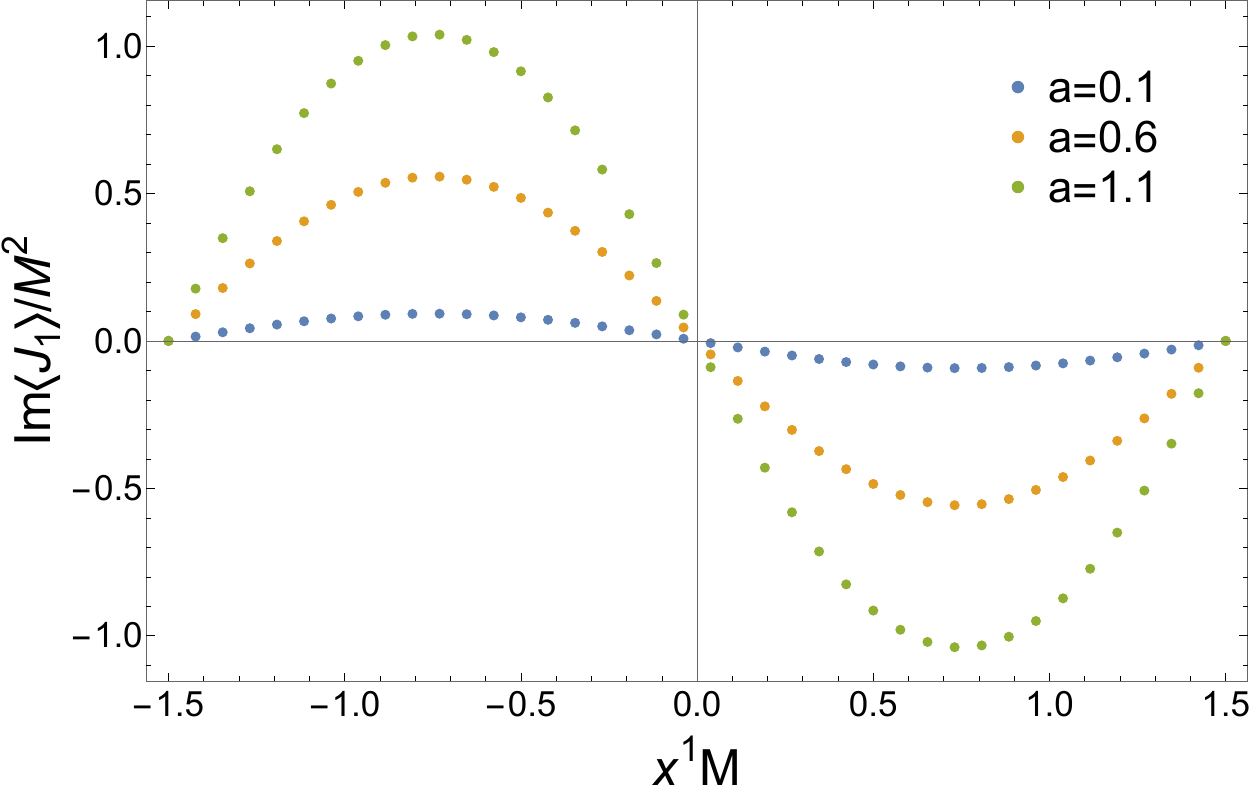}
    \caption{$T/M=0.5$}
    \label{fig:Current_PTbroken_T0d5}
\end{subfigure}
\hfill
\begin{subfigure}{.49\textwidth}
    \includegraphics[width=\textwidth]{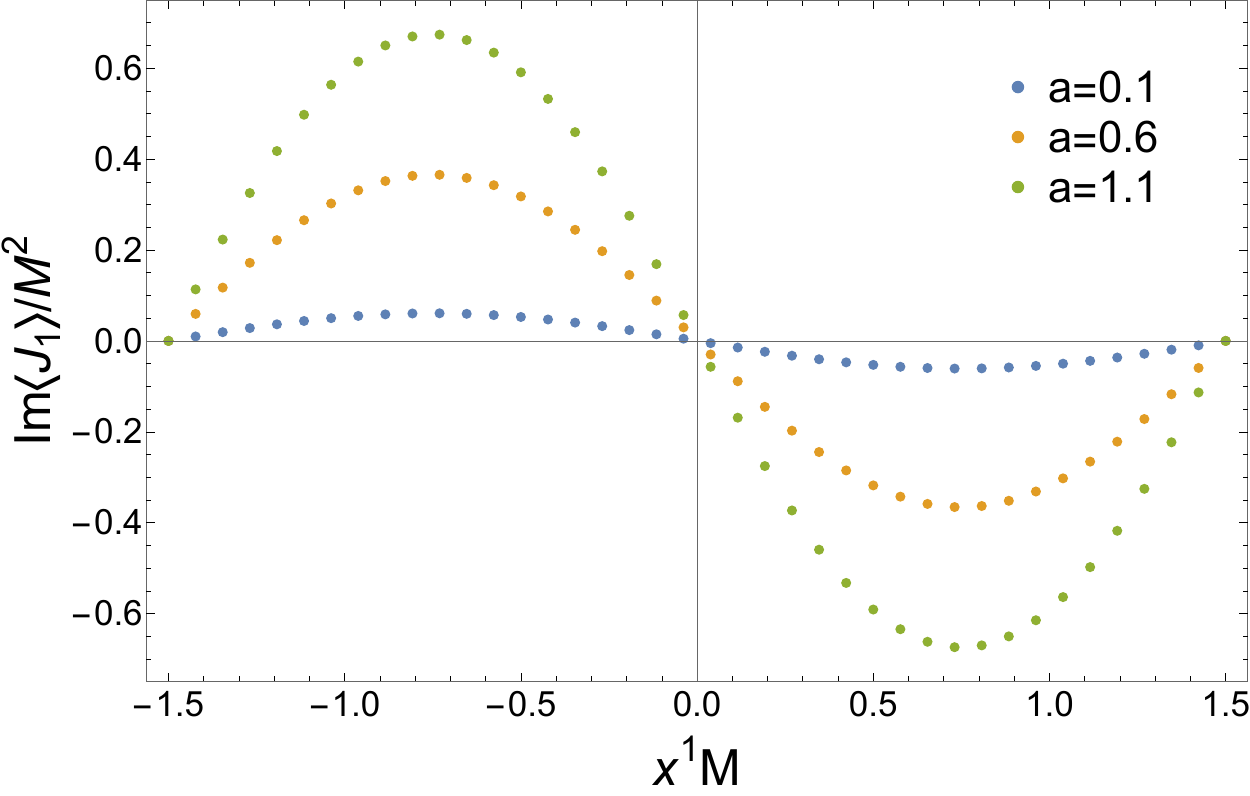}
    \caption{$T/M=1$}
    \label{fig:Current_PTbroken_T1d0}
\end{subfigure}
\caption{Imaginary part of the expectation value of the current $\expval{J_1}$ in a non-Hermitian lattice with $LM=3$. The real part is zero for any $\{a,T/M\}$. Figure taken from \cite{Arean:2024lzz}.}
\label{fig:Current_PTbroken}
\end{figure}

\begin{figure}
\captionsetup[subfigure]{justification=centering}
\centering
\begin{subfigure}{.49\textwidth}
    \includegraphics[width=\textwidth]{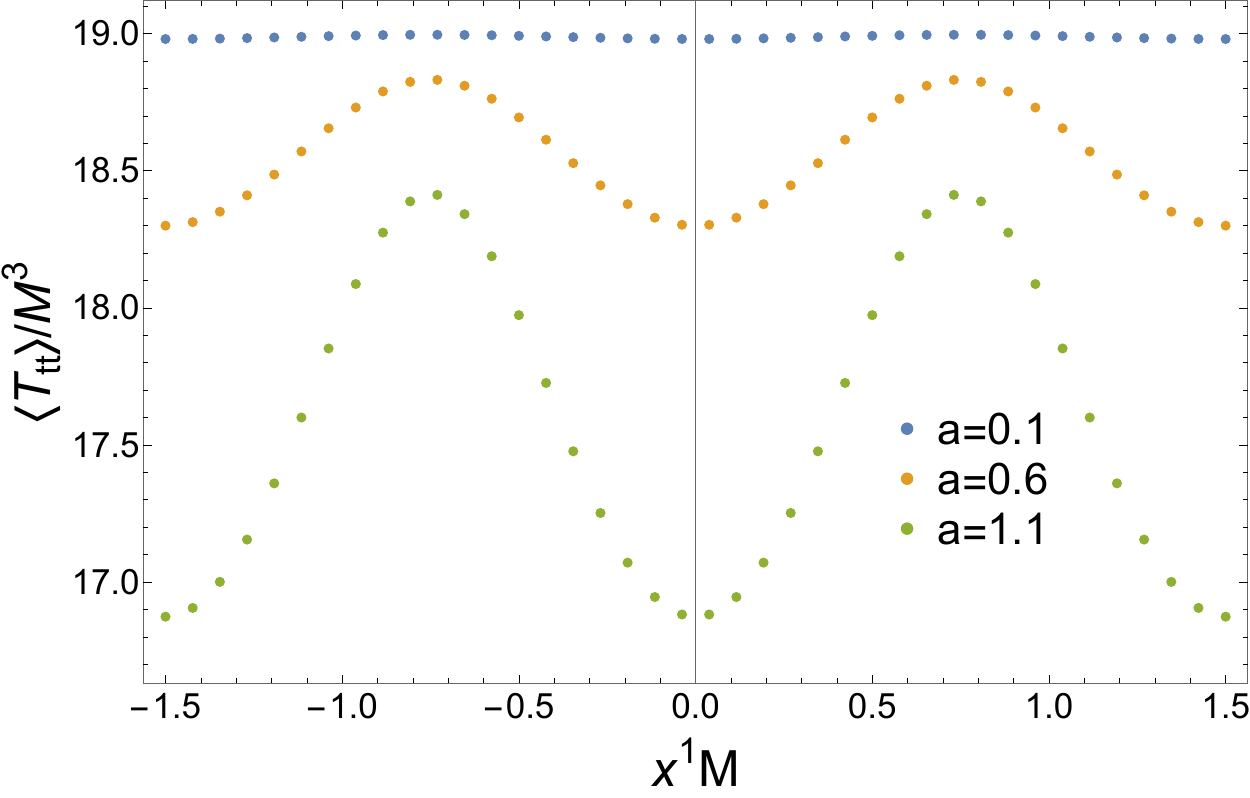}
    \caption{$T/M=0.5$}
    \label{fig:Ttt_PTbroken_T0d5}
\end{subfigure}
\hfill
\begin{subfigure}{.49\textwidth}
    \includegraphics[width=\textwidth]{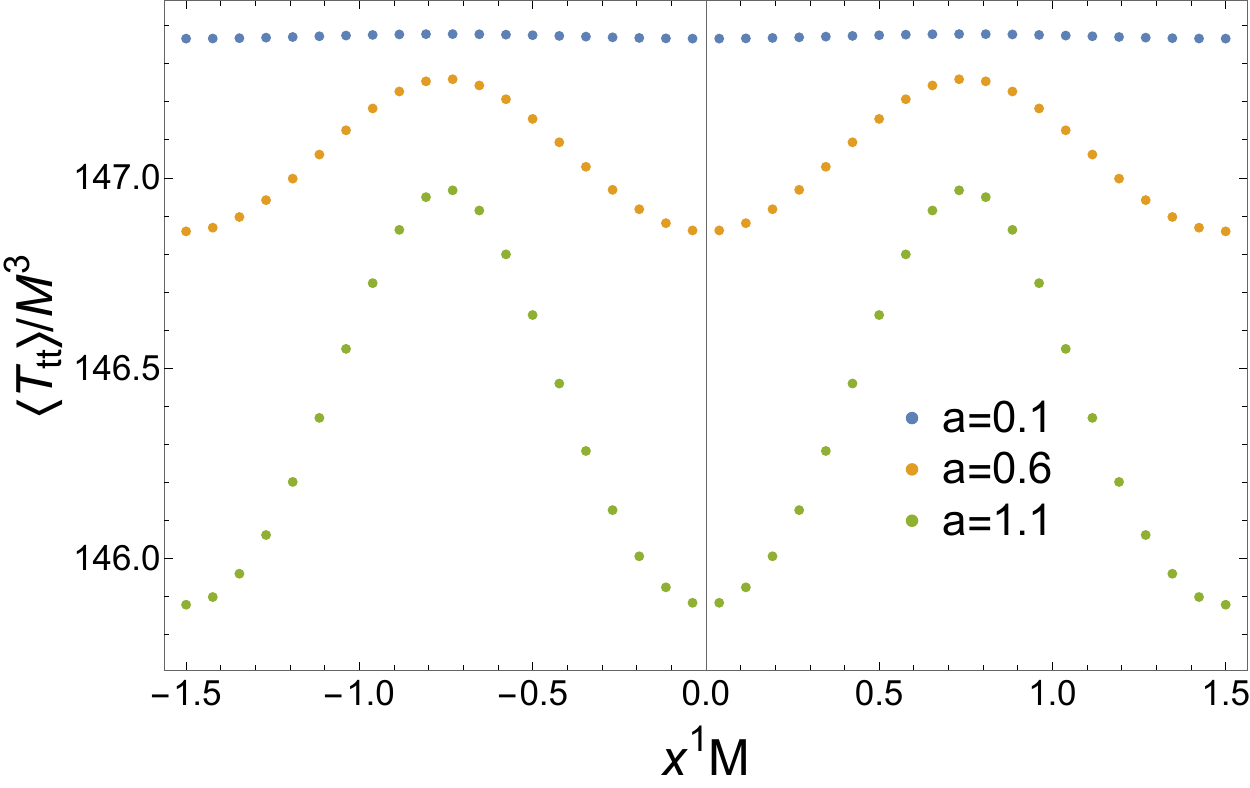}
    \caption{$T/M=1$}
    \label{fig:Ttt_PTbroken_T1d0}
\end{subfigure}
\caption{Expectation value of the energy density $\expval{T_{tt}}$ in a Non-Hermitian lattice with $LM=3$. Figure taken from \cite{Arean:2024lzz}.}
\label{fig:Ttt_PTbroken}
\end{figure}

The expectation values of the operators $\expval{{\mathcal{O}}}$, $\expval{\bar{\mathcal{O}}}$; the current $\expval{J_1}$ and the energy density $\expval{T_{tt}}$ for the non-Hermitian lattice are plotted in figures \ref{fig:Os_PTbroken}-\ref{fig:Ttt_PTbroken}. Remarkably these solutions have a real geometry and present a purely imaginary $U(1)$ current previously absent in the homogeneous setup. Most notably this current is an odd function of $x^1$ and thus spontaneously breaks $\mathcal{PT}$
\begin{equation}\label{eq:J_underPT}
    \expval{J_1(x)}\xrightarrow{\mathcal{PT}}\expval{J_1(\mathcal{PT}x)}=-\expval{J_1(x)}\neq\expval{J_1(x)}\,,
\end{equation}
indicating that this model cannot be mapped to a purely Hermitian theory via the Dyson map \eqref{eq:Dyson_Map}. Hence, inhomogeneities allow one to construct $\mathcal{PT}$-breaking solutions that do not have complex geometries. This is a generic feature of these solutions that holds for $a\neq0$. For $|a|>1$ sufficiently large at any fixed $T/M$ and $LM$ one finds that real solutions no longer exist. This resembles the observation of \cite{Arean:2019pom} for the homogeneous solutions of phase II. Hence it seems likely that complex solutions will dominate in some regions of the parameter space with $|a|$ large.

\begin{figure}
\centering
\includegraphics[width=0.6\textwidth]{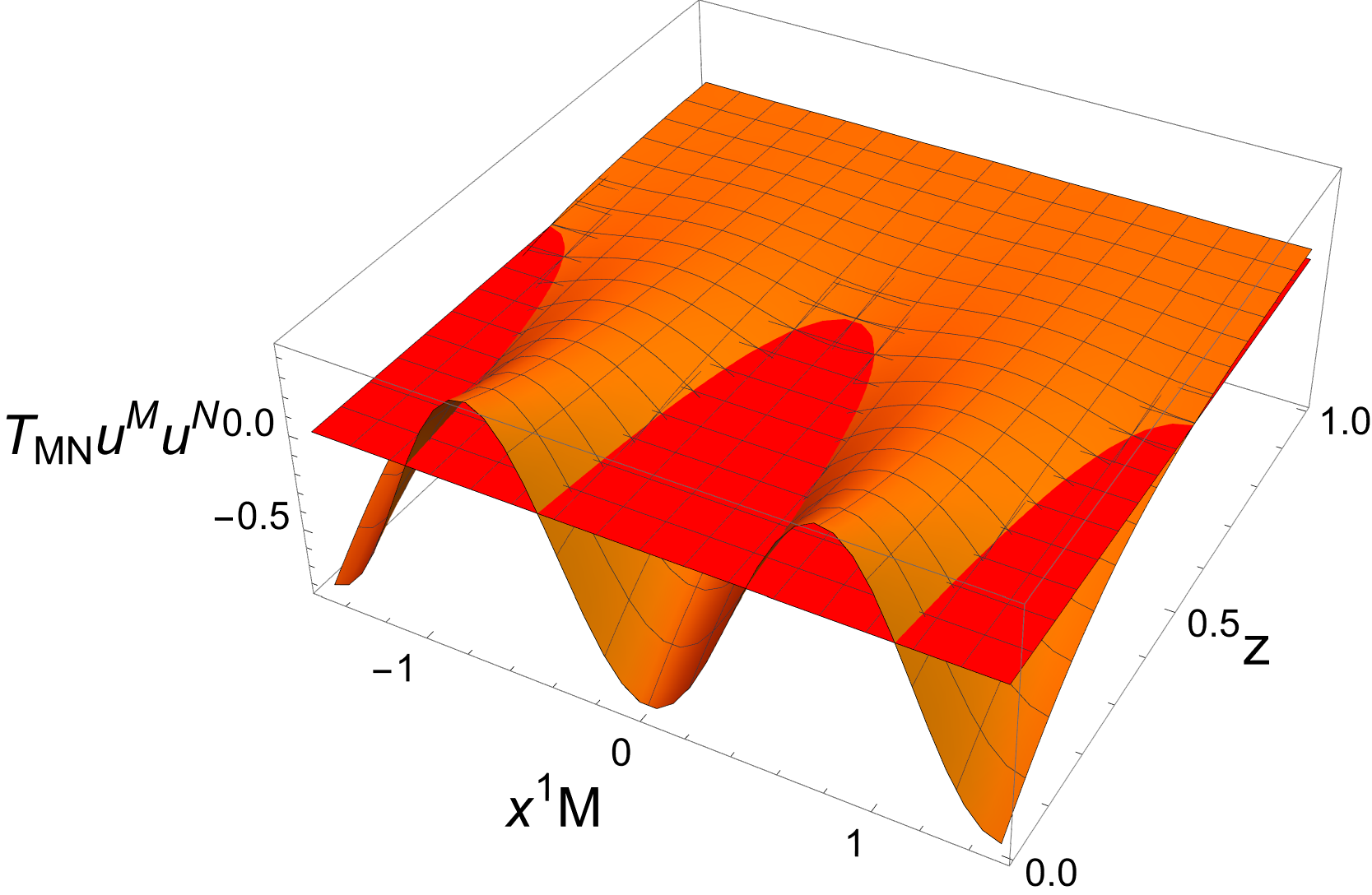}
\caption{Violation of the NEC for a stable non-Hermitian lattice with $T/M=0.5$ $LM=3$ and $a=1.7$. The orange surface denotes $T_{MN}u^Mu^N$ and the red one is the surface $T_{MN}u^Mu^N=0$ for reference. Here $u$ is an infalling null geodesic. Figure taken from \cite{Arean:2024lzz}.}
\label{fig:NEC_NHlattice}
\end{figure}

It is also worth mentioning that, unlike the homogeneous setup, stable $\mathcal{PT}$-breaking solutions were found even for setups with $|a|>1$ locally satisfying $|\eta|>1$ (which was the onset of instability for the homogeneous case). However, the solutions with $|a|>1$ do violate the NEC close to the AdS boundary (see figure \ref{fig:NEC_NHlattice}) although it is preserved near the event horizon. 

For solutions with $|a|>1$, a remarkable implication of the violation of the NEC near the AdS-boundary is that the standard $a$-function \cite{Freedman:1999gp,Myers:2010xs,Myers:2010tj} locally increases towards the IR
in the region where $|\eta(x^1)|>1$ . The $a$-function encodes the number of degrees of freedom along the renormalization group flow of the dual QFT. Hence these solutions are potentially problematic as the number of degrees of freedom seemingly increases as one integrates out from the UV to the IR. However, the failure of the standard a-function to be monotonically decreasing alone is not enough to conclusively prove this. To ensure that indeed the number of degrees of freedom increases, one would need to prove the non-existence of a monotonically decreasing function behaving as the central charge for $z\rightarrow0$.

\begin{figure}
\centering
\begin{subfigure}{.49\textwidth}
    \includegraphics[width=\textwidth]{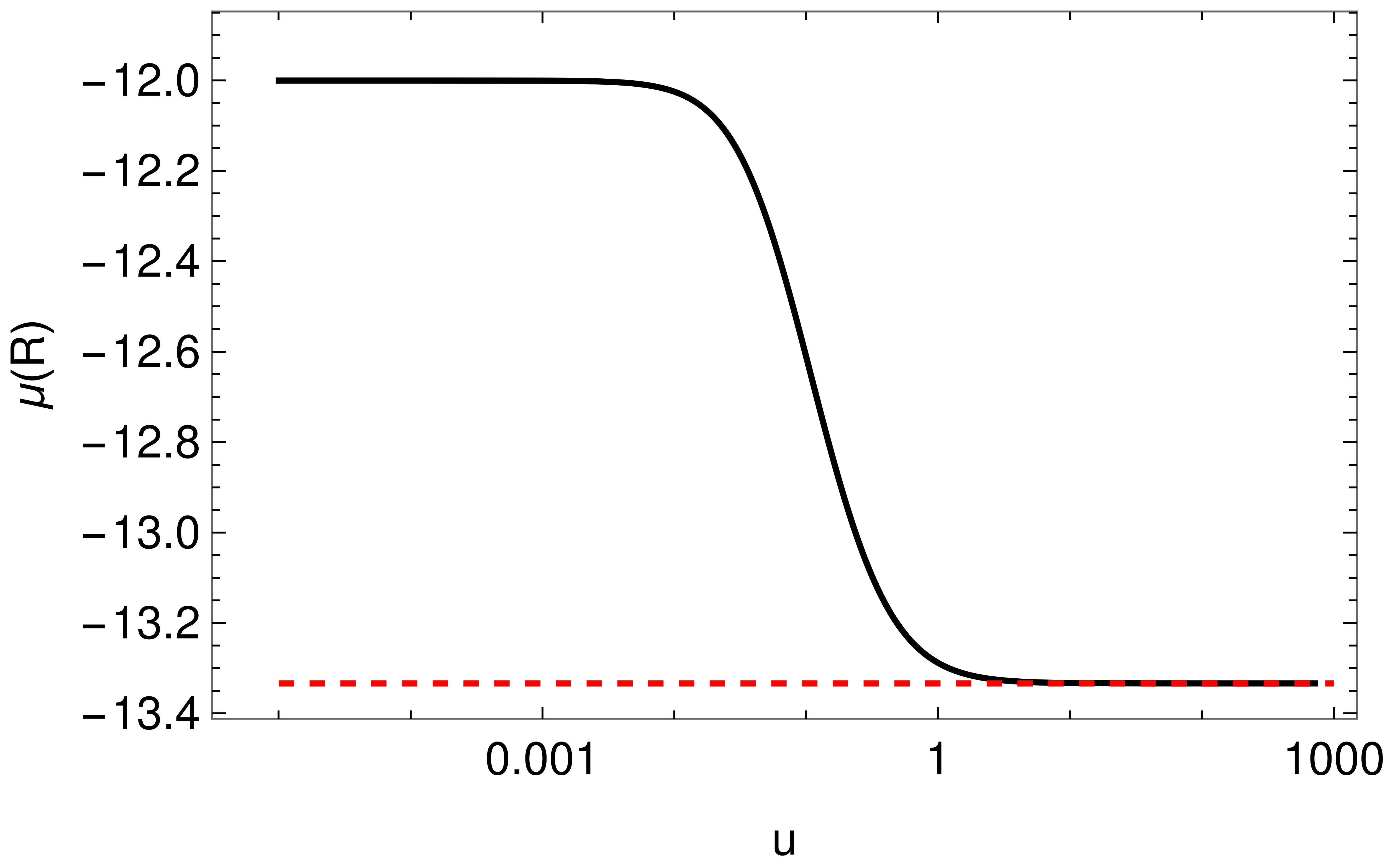}
    \label{fig:nhlattRicciMean}
\end{subfigure}
\hfill
\begin{subfigure}{.49\textwidth}
    \includegraphics[width=\textwidth]{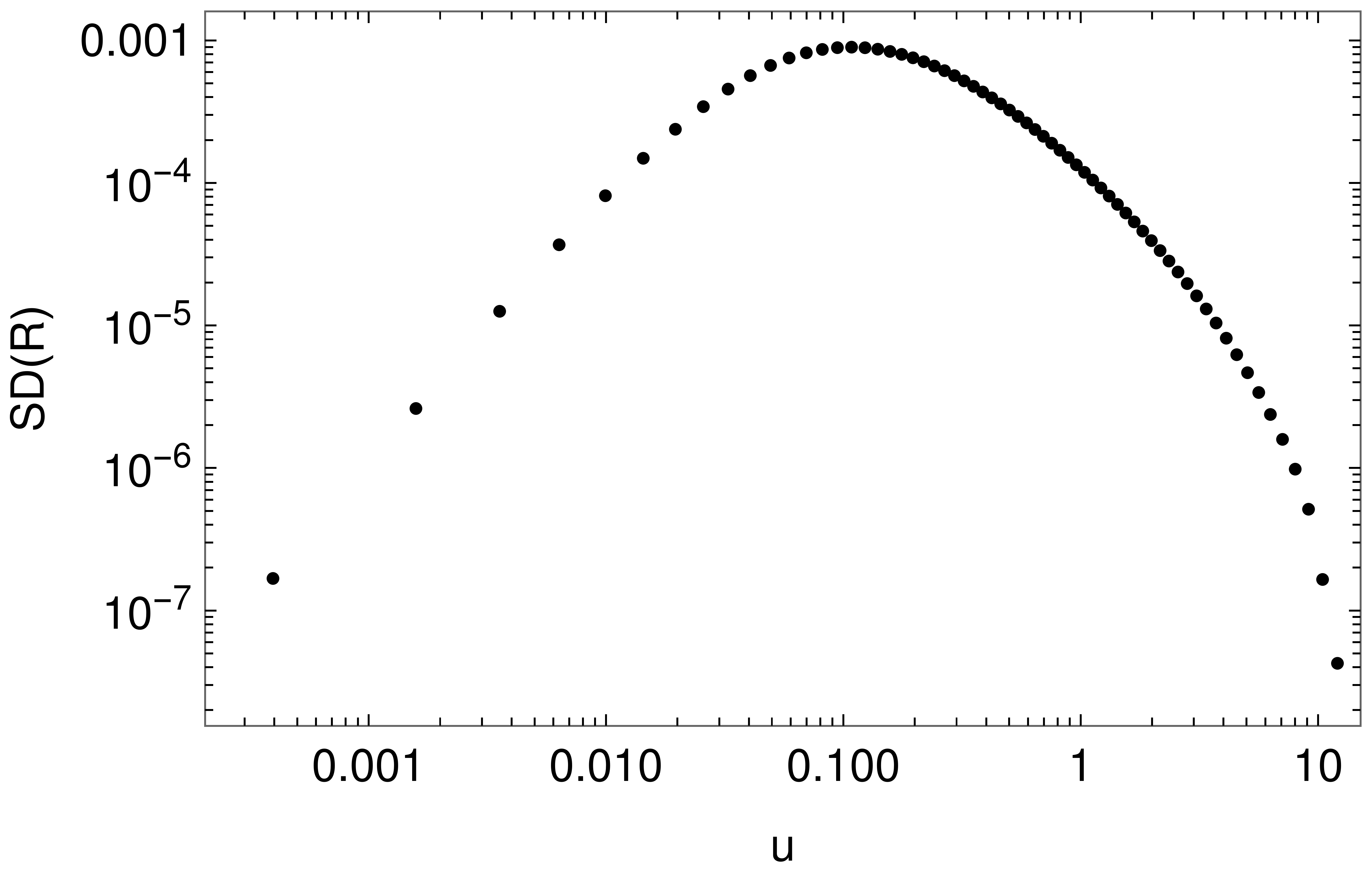}
    \label{fig:nhlattRicciSD}
\end{subfigure}
\caption{Left: spatial average of the Ricci scalar $R$ at zero temperature.
The red dashed line shows the value $R=-12-4v$
corresponding to the IR fixed point~\eqref{eq:NHHermitianIR}. Right: standard deviation of $R$ at zero temperature. The radial coordinate $u$ is defined as $u=z/(1-z)$.}
\label{fig:nhlattricci}
\end{figure}

\begin{figure}
\centering
\begin{subfigure}{.49\textwidth}
    \includegraphics[width=\textwidth]{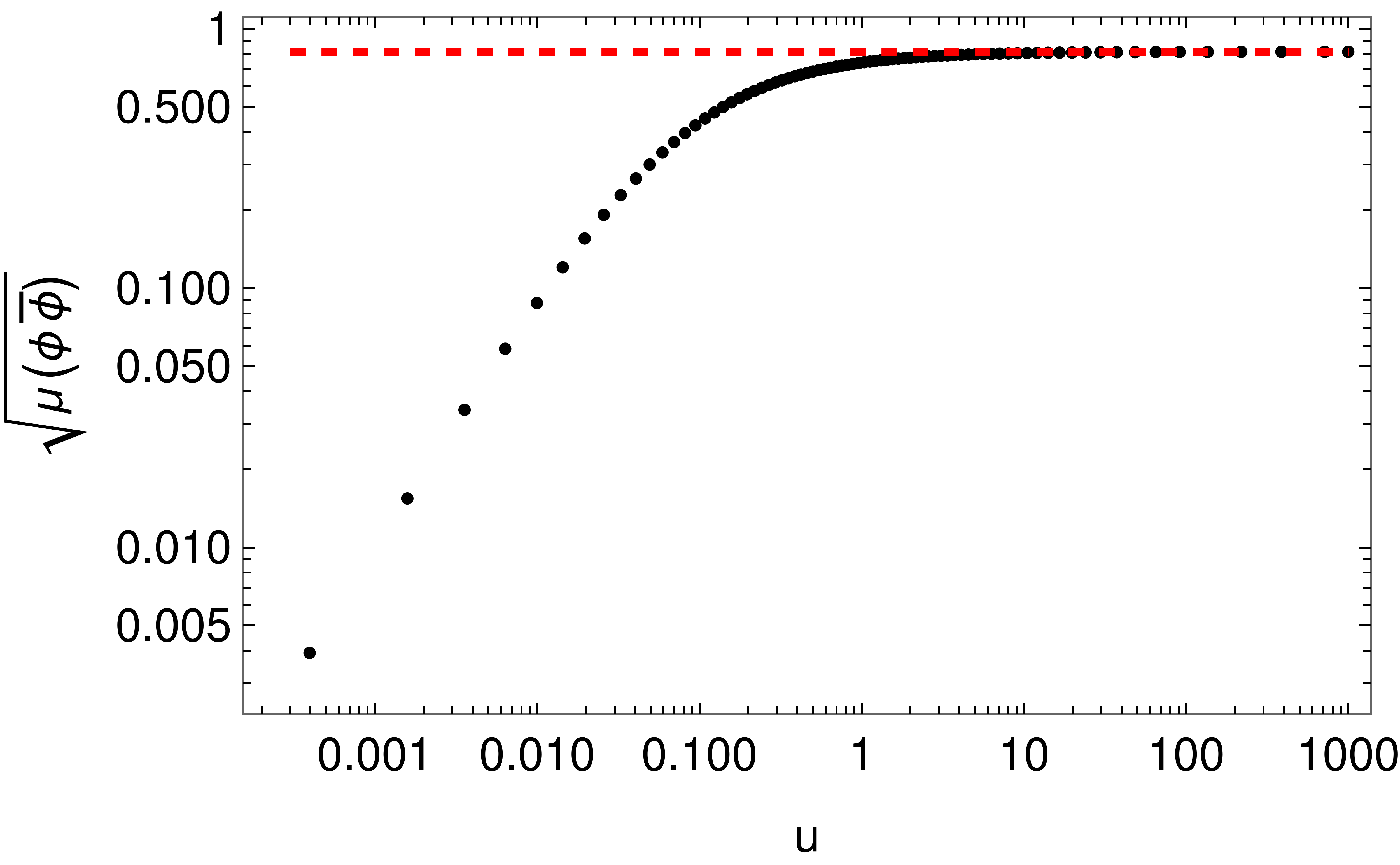}
    \label{fig:nhlatfibarfimean}
\end{subfigure}
\hfill
\begin{subfigure}{.49\textwidth}
    \includegraphics[width=\textwidth]{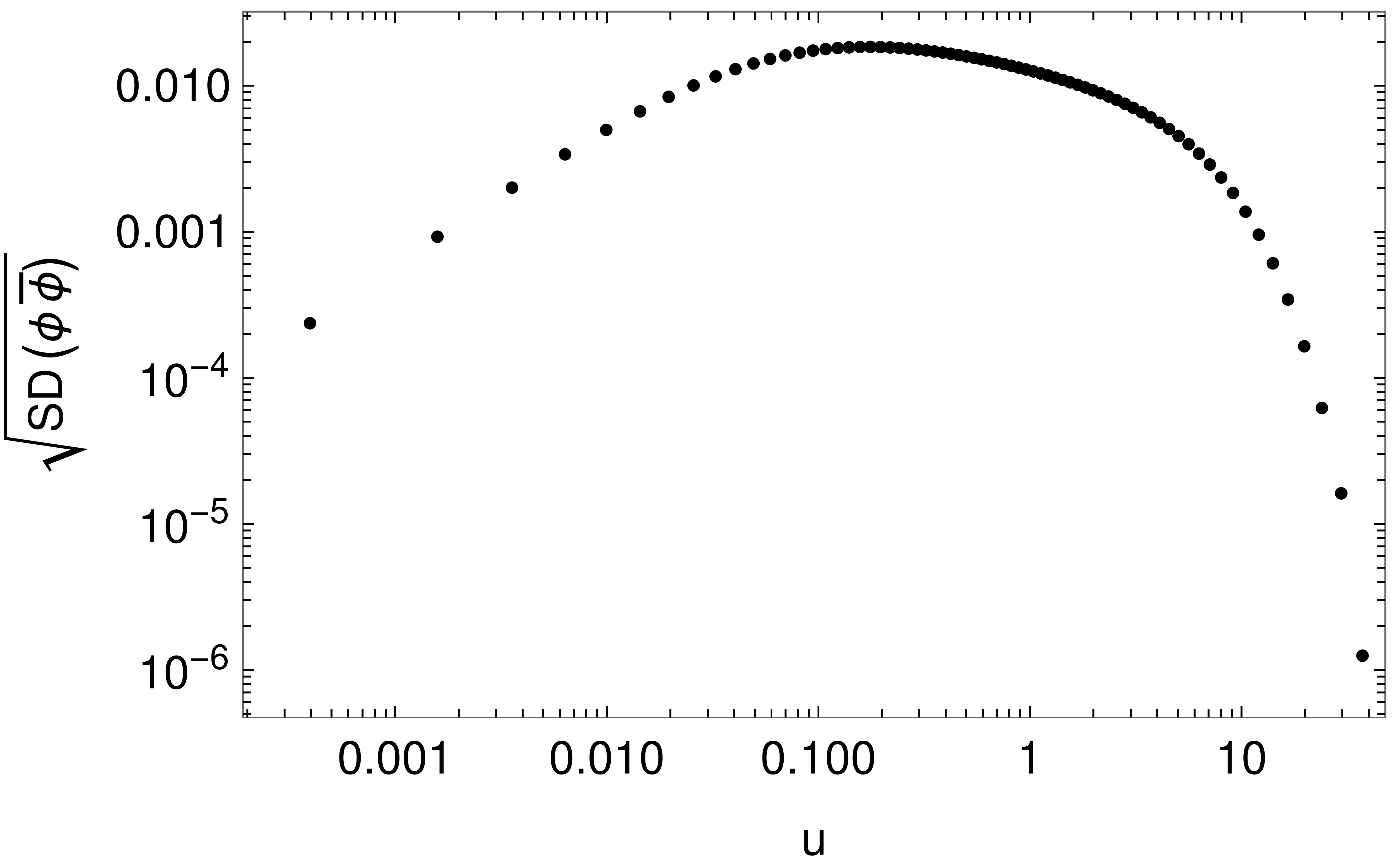}
    \label{fig:nhlatfibarfisd}
\end{subfigure}
\caption{Left: square root of the spatial average of $\phi\bar\phi$ at zero temperature. The red dashed line depicts the value $\sqrt{2/v}$
corresponding to the IR fixed point~\eqref{eq:NHHermitianIR}. Right: square root of the standard deviation of $\phi\bar\phi$ at zero temperature. As above $u=z/(1-z)$}
\label{fig:nhlatfibarfi}
\end{figure}

In \cite{Arean:2024lzz} the authors also studied the zero-temperature limit\footnote{Notice that to find the zero-temperature solutions one needs to modify the ansatz \eqref{eq:Ansatz_Poincare} as indicated in \cite{Arean:2024lzz}.} of the non-Hermitian lattice for $|a|<1$. They found that the zero temperature IR geometry corresponds to that of a Hermitian conformal fixed point \cite{Gubser:2008wz}\footnote{For the zero-temperature solutions the charge is set to $q=2$. As noted in \cite{Gubser:2009cg}, this choice ensures the existence of a conformal IR fixed point for our choice of parameters.}
\begin{align}\label{eq:HermitianIR}
    ds^2=&\frac{3v }{(1+3v)z^2} \left(-dt^2+(dx^1)^2+(dx^2)^2+dz^2\right)\,,\nonumber\\
    &\phi=\bar\phi=\sqrt{2\over v}\,,\quad A_1=0\,,
\end{align}
rotated under a complexified $U(1)$. This is illustrated in figures \ref{fig:nhlattricci} and \ref{fig:nhlatfibarfi} where we plot the standard deviation and the mean value of the Ricci scalar $R$ and the $U(1)$ invariant $\phi\bar{\phi}$;
observing that they tend to the values at the conformal fixed point $R=-12-4/v$ and $\sqrt{\phi\bar{\phi}}=\sqrt{2/v}$. 

In more detail, the authors found that the solution near the IR had the following form
\begin{align}\label{eq:NHHermitianIR}
    ds^2&=\frac{3v }{(1+3v)z^2} \left(-dt^2+(dx^1)^2+(dx^2)^2+dz^2\right)\,,\nonumber\\
\phi&=S \sqrt{2\over v}\,,\quad\bar\phi=S^{-1} \sqrt{2\over v}\,,\quad  A_1=-\frac{i´}{q}S^{-1}\partial_{1}S\,,
\end{align}
with $S$ a $x^1$-dependent complexified $U(1)$ rotation given by
\begin{equation}\label{eq:sdeflowT}
    S=\sqrt{1-\tilde \eta\over 1+\tilde\eta}\,,
\end{equation}
where numerically it was observed that the function $\tilde{\eta}$ satisfied
\begin{equation}
{\tilde \eta\over \sqrt{1-\tilde\eta^2}}=\eta\,.   
\end{equation}

In conclusion, the IR geometry restores $\mathcal{PT}$ as a Dyson map of the form \eqref{eq:Dyson_Map} connects it to the Hermitian fixed point \eqref{eq:HermitianIR} which is $\mathcal{PT}$-unbroken. Hence the non-Hermitian lattices flow from a $\mathcal{PT}$-broken UV to a $\mathcal{PT}$-unbroken theory in the IR. Quite remarkably, the IR $\mathcal{PT}$-symmetry restoration observed in this strongly-coupled model, was also found in the perturbative regime in \cite{Chernodub:2021waz,Chernodub:2024lkr}. 
There, the authors found that small non-Hermitian deformations to the mass matrix of a doublet model lead to a theory interpolating between a $\mathcal{PT}$-broken, non-unitary UV and a $\mathcal{PT}$-symmetric, unitary IR. 
This suggests that this phenomenon might be a generic property of inhomogeneous non-Hermitian systems although further studies are needed to determine if this is indeed the case. 

\section{Conclusions}\label{sect:Conclusions}

Non-Hermitian QFTs are very interesting due to their novel phenomenology absent in their Hermitian counterparts. They allow us to describe open systems and by tuning the matter flow we can achieve stationary solutions where unitary evolution can be realized. This is directly related to the existence of an unbroken antilinear $Z_2$-symmetry; typically taken to be $\mathcal{PT}$ which guarantees that the energy spectrum is real. 

Much of the study of non-Hermitian QFTs has been restricted to the perturbative regime for its simplicity. However, in order to deepen our understanding of these theories, it is interesting to extend such studies to the strong coupling regime. In this context, holographic non-Hermitian QFTs provide an interesting path as they offer a controlled non-perturbative setup. 

In this work we have collected the key developments in the study/field of holographic non-Hermitian field theories. 
We have reviewed the construction of a gravity dual to a CFT where a non-Hermitian deformation is introduced through the sources of a complex scalar operator~\cite{Arean:2019pom}. The model is $\mathcal{PT}$-symmetric and features a rich phase diagram with phases where $\mathcal{PT}$ is spontaneously broken and phases where it is not~\cite{Xian:2023zgu}. We have shown how the Dyson map, that maps $\mathcal{PT}$-symmetric phases to a Hermitian description is implemented in the gravity dual.
We next considered configurations where the non-Hermitian deformation is spacetime-dependent~\cite{Morales-Tejera:2022hyq}.
We distinguished two cases: quenches \cite{Morales-Tejera:2022hyq} and lattices \cite{Arean:2024lzz}. In the former, the non-Hermitian deformation is time-dependent while in the latter is space-dependent. For quenches from a Hermitian theory to a non-Hermitian theory, it was observed that the apparent horizon of the black brane shrinks, indicating that the dual field theory decreases its entropy. On the other hand, for the lattices it was observed an spontaneous breaking of $\mathcal{PT}$ with real geometries. Furthermore this symmetry was restored in the IR, indicating that the model presents a $\mathcal{PT}$-restoring RG flow.

\section{Future directions}\label{sect:FutureDirections}
We conclude this review with a list of
possible future research directions in the context of non-Hermitian holography:

\begin{itemize}
    \item \textbf{Construct holographic models with non-Hermitian $SU(2)$ sources.} The existing literature of non-Hermitian models in holography has focused on the case where the non-Hermiticity arises from the sources of a scalar operator $\mathcal{O}$ charged under $U(1)$. It would be interesting to generalize these ideas to models with non-Hermitian sources for operators charged under $SU(2)$. In the context of spacetime-dependent sources, one could think of this setup as a strongly-coupled generalization of the models of \cite{Chernodub:2021waz,Chernodub:2024lkr}.

    \item \textbf{Fermi surfaces in non-Hermitian holography}. In \cite{Zhen:2015} non-Hermitian Dirac cones were studied and experimentally realized in an optical setting. It would be interesting to study the effect of the non-Hermiticity on the Fermi surface of the holographic model with constant sources to compare with those results. Fermi surfaces correspond to the poles of fermionic correlators which are dual to quasinormal modes of probe fermions in the bulk.

    \item \textbf{Non-hermitian cooling}.
    We have highlighted the interesting phenomenon that non-Hermitian holographic quantum quenches can violate the null energy condition on the Horizon of an asymptotically AdS black hole. This makes the (apparent) horizon shrink and leads effectively to a lowering of the Hawking temperature. Up to our knowledge this effect of non-Hermitian cooling has not been investigated yet at weak coupling. It would be interesting to study it in non-holographic models. 
    Since $\mathcal{PT}$-symmetric non-Hermitian quantum systems can be realized in experiment it is also tempting to contemplate if such a non-Hermitian quantum cooling could be observed in the laboratory. 

    \item \textbf{Phase diagram of non-Hermitian lattices}. In \cite{Xian:2023zgu} the full phase diagram was constructed for the model with constant sources. It would be interesting to generalize such phase diagram to the case of inhomogeneous sources. In particular one could study the existence of the solutions with complex geometries and whether such solutions are thermodynamically preferred.

    \item \textbf{Universal nature of IR $\mathcal{PT}$-restoration in inhomogeneous models}. Perturbatively, it was observed in \cite{Chernodub:2021waz} that $\mathcal{PT}$ symmetry is restored in inhomogeneous non-Hermitian models with a $\mathcal{PT}$-broken UV. This same phenomenon was also observed in the non-Hermitian lattice of~\cite{Arean:2024lzz} reviewed here. It would be interesting to explore whether this IR $\mathcal{PT}$-symmetry restoration is a generic property of inhomogeneous non-Hermitian field theories. 

    \item \textbf{Non-Hermitian deformations around quantum critical points}. In~\cite{Ashida_2017} it was observed that a non-Hermitian lattice can qualitatively modify the dynamics in the vicinity of a quantum critical point. Via holography one can explore these effects on strongly coupled systems.

\end{itemize}

\vspace{6pt} 

\authorcontributions{All authors contributed equally.}

\funding{This work is supported through the grants CEX2020-001007-S and PID2021-123017NB-100, PID2021-127726NB-I00 funded by MCIN/AEI/10.13039/501100011033 and by ERDF ``A way of making Europe''. The work of D.G.F. is supported by FPI grant PRE2022-101810.}

\acknowledgments{We thank S. Morales-Tejera for his help with the figures of section \ref{ssect:nhquenches} and Z.Y. Xian, D. Rodríguez Fernández, Z. Chen, Y. Liu and R. Meyer for answering our questions concerning their paper \cite{Xian:2023zgu}.}

\begin{adjustwidth}{-\extralength}{0cm}

\reftitle{References}

\bibliography{biblio.bib}

\end{adjustwidth}

\end{document}